\documentclass[12pt,thmsa]{article}
\usepackage{amssymb}

\usepackage{sw20rui}

\input{tcilatex}
\usepackage[T1]{fontenc}
\begin{document}

\title{The Universal Cut Function and Type II Metrics }
\date{Nov. 29, 2006 }
\author{Carlos Kozameh$^{1}$ \and E.T. Newman$^{2}$ \and J.G. Santiago-Santiago$^{3}$
\and Gilberto Silva-Ortigoza$^{3}$ \\
%EndAName
$^{1}$FaMaF, Univ. of Cordoba, \\
Cordoba, Argentina\\
$^{2}$Dept of Physics and Astronomy, \\
Univ. of Pittsburgh, \\
Pittsburgh, PA 15260, USA\\
$^{3}$Facultad de Ciencias F\'{\i }sico Matem\'{a}ticas \\
de la Universidad Aut\'{o}noma de Puebla, \\
Apartado Postal 1152, 72001,\\
Puebla, Pue., M\'{e}xico}
\maketitle

\begin{abstract}
In analogy with classical electromagnetic theory, where one determines the
total charge and both electric and magnetic multipole moments of a source
from certain surface integrals of the asymptotic (or far) fields, it has
been known for many years - from the work of Hermann Bondi - that energy and
momentum of gravitational sources could be determined by similar integrals
of the asymptotic Weyl tensor. Recently we observed that there were certain
overlooked structures, \textit{defined at future null infinity,} that
allowed one to determine (or define) further properties of both
electromagnetic and gravitating sources. These structures, families of 
\textit{complex} `slices' or `cuts' of Penrose's $\frak{I}^{+}$, are
referred to as Universal Cut Functions, (UCF). In particular, one can define
from these structures a (complex) center of mass (and center of charge) and
its equations of motion - with rather surprising consequences. It appears as
if these asymptotic structures contain in their imaginary part, a well
defined total spin-angular momentum of the source. We apply these ideas to
the type II algebraically special metrics, both twisting and twist-free.
\end{abstract}

\section{Introduction}

In the present work we return to the study of an old subject\cite
{Bondi,Sachs,NP1,NU}, namely properties of asymptotically flat Einstein
space-times, both vacuum and with a Maxwell field. More specifically we
explore some previously overlooked properties of such space-times. The point
of view is however rather new.

The basic idea is that for each asymptotically flat vacuum solution there is
a certain `structure', a unique function on null infinity, $\frak{I}^{+},$
that has been known in various contexts, but whose several unusual
properties have largely been overlooked. This function, which we shall refer
to as the Universal Cut Function (UCF), carries virtually all the asymptotic
information, as well as much of the physical content of its associated
solution. In addition, it unifies several apparently disparate ideas. In the
case of the Einstein-Maxwell fields, in general there are two such
functions, one for GR and one for the Maxwell field - however we will
consider only the degenerate case where the two function are the same.

We begin with the conventional view of $\frak{I}^{+};$ it is a
three-dimensional manifold, $S^{2}$x$R,$ coordinatized by the standard Bondi
coordinates $(u_{r}$,$\zeta ,\overline{\zeta }$), with complex stereographic
coordinates ($\zeta ,\overline{\zeta }$) on the $S^{2}$ and $u_{r}$ on the $%
R $ part. (The subscript `$r$' denotes the retarded time.) We part from the
conventional point of view by allowing the complexification of $\frak{I}%
^{+}, $ i.e., we consider $\frak{I}_{C}^{+}$ with $u_{r}$ now allowed to
take complex values close to the real and ($\zeta ,\overline{\zeta }$) going
to ($\zeta ,\widetilde{\zeta }$) with $\widetilde{\zeta }$ close to $%
\overline{\zeta }.$ It should be emphasized that this complexification of $%
\frak{I}^{+} $ is a formal device that has many interesting consequences.
Nevertheless we are dealing with real space-times and real solutions to the
Einstein equations.

A significant notational problem should be addressed now. The time variable $%
u_{r}$ used above is, in flat space the \textit{retarded} time $t-r.$ Many
years ago\cite{ENP}, in order to keep the symmetry of many of the equations
in the spin-coefficient formalism this $u_{r}$ was replaced by $u,$ via 
\begin{equation}
u_{r}=\sqrt{2}u.  \label{sqrt2}
\end{equation}
Later to see the physical consequences of our results we restore the
retarded time $u_{r}$ and, in addition, explicitly display the velocity of
light, $c.$

In Sec. II we review some background material. Much of it, though in the
literature, is still quite unfamiliar to most readers. To help overcome this
unfamiliarity, we present in Sec. III a detailed discussion, in the
simplified case of flat space-times$,$ of the ideas to be used later in the
asymptotically flat cases. Sec. IV is devoted to a discussion of these ideas
for general asymptotically flat space-times while Sec. V is used to
specialize the discussion to the algebraically special type II
Einstein-Maxwell metrics. In Sec. V and the remaining sections, we first see
how the new ideas interact with the classical Bondi energy-momentum
four-vector, giving it new properties and even fresh predictions. Then we
see how it interacts with the Bondi energy-momentum loss equation, giving it
surprising new interpretations that include equations of motion for the
center of mass motion, even containing radiation reaction forces. It even
strongly suggests a general relativistic definition of spin-angular
momentum. Finally in Sec. VIII, the conclusion, we discuss how in principle
one could try to observe the predictions described earlier.

\section{Background}

The UCF, which had been introduced earlier in different contexts\cite
{NU,Posadas,RP2}, is described as follows: It is to be a one-complex
parameter family of `cuts' - complex 2-surfaces - of $\frak{I}_{C}^{+}$
written as 
\begin{equation}
u=X(\tau ,\zeta ,\widetilde{\zeta }),  \label{X}
\end{equation}
with (in general) a complex $\tau $ parametrizing the family. [In important
special cases, we could have a \textit{real} one parameter family of real
cuts of $\frak{I}^{+},$ so that $u=X(\tau ,\zeta ,\overline{\zeta })$ was a
real function of a real $\tau .$ In fact, in this case, Eq.(\ref{X}) could
be treated as a coordinate transformation from $u$ to $\tau .$ The real
coordinates ($\tau ,\zeta ,\overline{\zeta })$ have been referred to NU
coordinates.] We also assume that Eq.(\ref{X}) can be inverted, i.e., solved
for 
\begin{equation}
\tau =T(u,\zeta ,\widetilde{\zeta }).  \label{T}
\end{equation}
For real $u$ and $\widetilde{\zeta }=\overline{\zeta }$ this obviously
defines a complex $\tau $ such that the $u$ in Eq.(\ref{X}) is real.
\{Aside: Later we will see that $T$ for real $u$ and $\widetilde{\zeta }=%
\overline{\zeta }$ is a CR function for a CR structure obtained from $X(\tau
,\zeta ,\widetilde{\zeta }).$\}

From $X(\tau ,\zeta ,\widetilde{\zeta })$ we define two functions, $V$ and $
L,$ (that have been extensively used in GR in different contexts, with
different coordinate systems\cite{NT}, by 
\begin{eqnarray}
V &=&\partial _{\tau }X\equiv X^{\prime }(\tau ,\zeta ,\widetilde{\zeta }),
\label{V} \\
L &=&\text{\dh }_{(\tau )}X(\tau ,\zeta ,\widetilde{\zeta }),  \label{L}
\end{eqnarray}
with \dh $_{(\tau )}$ meaning the edth operator with $\tau $ held constant.
Here, these functions have a common origin in the UCF. An important point is
that, in both $V$ and $L,$ the $\tau $ could be eliminated via Eq.(\ref{T})
and they could be treated as functions on the real $\frak{I}^{+}.$ Both have
simple geometric meanings: $V=X^{\prime }(\tau ,\zeta ,\widetilde{\zeta })$
is the `$u$' separation between neighboring constant $\tau $ cuts while $%
L(u,\zeta ,\overline{\zeta })$ is a stereographic `angle' field on real $%
\frak{I}^{+},$ on the sphere of past null directions, that determines the
directions of an asymptotically shear-free null geodesic congruence going
into the interior.

In the case of algebraically special type II space-times, $V(\tau ,\zeta ,%
\overline{\zeta })$ is the basic variable in the Robinson-Trautman metrics%
\cite{RT} with real $\tau ,$ while $L(u,\zeta ,\overline{\zeta })$ is the
basic variable for the twisting type II metrics.

\begin{itemize}
\item  
\begin{center}
Note that we use $(^{\prime })=\partial _{\tau }$ and $(^{\cdot })=\partial
_{u}$
\end{center}
\end{itemize}

Of primary importance to us is the fact that $X$ and hence $L$ and $V$ are
uniquely determined (up to gauge transformations - BMS transformations) in
any given (non-flat) asymptotically flat space-time. They are determined by
the appropriate choice of a \textit{shear-free} (for flat or algebraically
special space-times) or an \textit{asymptotically shear-free} (for
asymptotically flat space-time) \textit{null geodesic congruence. }

\textit{For flat space-time}, using the variables $(u,\zeta ,\widetilde{%
\zeta }),$ $X$ and $L$ are determined by first solving the equation for
shear-free null geodesic congruences\cite{A}, 
\begin{equation}
\text{\dh }_{(u)}L(u,\zeta ,\widetilde{\zeta })+LL^{{\large \cdot }}=0,
\label{ethL}
\end{equation}
for $L(u,\zeta ,\widetilde{\zeta })$ followed by solving the so-called CR
equation 
\begin{equation}
\text{\dh }_{(u)}T+L(u,\zeta ,\widetilde{\zeta })T^{\,{\large \cdot }}=0,
\label{ethT}
\end{equation}
to determine $\tau =T(u,\zeta ,\widetilde{\zeta }).$ Finally by inverting $%
T(u,\zeta ,\widetilde{\zeta }),$ we obtain $u=X(\tau ,\zeta ,\widetilde{%
\zeta }).$

\begin{itemize}
\item  \dh $_{(u)}$ is the eth operator with $u$ held constant.
\end{itemize}

An alternate and easier method to find $X$ and $L$ , is to immediately use
the variables ($\tau ,\zeta ,\widetilde{\zeta }),$ and solve the equation 
\begin{equation}
\text{\dh }_{(\tau )}^{2}X(\tau ,\zeta ,\widetilde{\zeta })=0,
\label{eth^2x}
\end{equation}
for $u=X(\tau ,\zeta ,\widetilde{\zeta })$ and then find $L(\tau ,\zeta ,%
\widetilde{\zeta })$ by 
\begin{equation}
L(\tau ,\zeta ,\widetilde{\zeta })=\text{\dh }_{(\tau )}X(\tau ,\zeta ,%
\widetilde{\zeta }),  \nonumber
\end{equation}
noticing that 
\[
\text{\dh }_{(\tau )}L(\tau ,\zeta ,\widetilde{\zeta })=0. 
\]

These results follow by implicit differentiation of Eq.(\ref{X}) using Eq.(%
\ref{ethL}). In the flat space case there is considerable freedom in the
choice of solution for $X$ and $L$ that is considerably narrowed by the
requirement of regularity of the function $L$ on $S^{2}.$ The geometric
meaning of `regularity' is that all the null geodesics are transverse to $%
\frak{I}^{+},$ i.e. none of the geodesics lie on $\frak{I}^{+}.$ The final
freedom, which follows from Eq.(\ref{eth^2x}), is that of a free complex
analytic word-line in complex Minkowski space. We return in more detail to
this issue in Sec. II.

\textit{For asymptotically flat space-times} with a given Bondi asymptotic
shear, $\sigma =\sigma (u,\zeta ,\widetilde{\zeta }),$ generalizing the flat
space treatment, the UCF is determined by first solving

\begin{equation}
\text{\dh }_{(u)}L(u,\zeta ,\widetilde{\zeta })+LL^{{\large \cdot }}=\sigma
(u,\zeta ,\widetilde{\zeta }),  \label{ethL*}
\end{equation}
for $L(u,\zeta ,\widetilde{\zeta }),$ followed by solving for $T$ the $CR$
equation

\begin{equation}
\text{\dh }_{(u)}T+L(u,\zeta ,\widetilde{\zeta })T^{\,{\large \cdot }}=0.
\label{ethT*}
\end{equation}
By inverting the solution, Eq.(\ref{T}), we obtain $X(\tau ,\zeta ,%
\widetilde{\zeta }).$ Again there is considerable freedom in the solution
which is first narrowed again using regularity on $S^{2},$ then made\textit{%
\ unique} by certain algebraic conditions on the Weyl tensor components\cite
{KNO,KNO2,FtPrints}.

An alternative method of finding the UCF from a given $\sigma (u,\zeta ,%
\widetilde{\zeta })$ is to solve the PDE, the so-called good-cut equation%
\cite{KNT,HNPT},

\begin{equation}
\text{\dh }^{2}Z(\zeta ,\widetilde{\zeta })=\sigma (Z,\zeta ,\widetilde{
\zeta }).  \label{GoodCutEq}
\end{equation}
The solutions, which are known\cite{KNT,HNPT} to depend on four complex
parameters, say $z^{a},$ can be written as 
\begin{equation}
Z=Z(z^{a},\zeta ,\widetilde{\zeta }).  \label{Z}
\end{equation}
When $z^{a}$ are taken as arbitrary complex analytic functions of $\tau ,$
i.e., a complex world-line in the parameter space,

\begin{equation}
z^{a}=\xi ^{a}(\tau ),  \label{world-line}
\end{equation}
we\textit{\ }have 
\begin{equation}
u=X(\tau ,\zeta ,\widetilde{\zeta })=Z(\xi ^{a}(\tau ),\zeta ,\widetilde{
\zeta }),  \label{Z=>X}
\end{equation}
which is \textit{almost, }but not quite\textit{,} the UCF.

At this point we have:

\begin{theorem}
\textit{Every transverse (regular) shear free or asymptotically shear-free
null geodesic congruence in an asymptotically flat space-time is generated
by a complex world-line in H-space.}
\end{theorem}

Finally an algebraic condition on the Weyl tensor components determines\cite
{KNO,KNO2,FtPrints,KN} a \textit{unique} complex world-line and thus yields
the UCF.

One knows from electrodynamics that many properties of the sources (charge
and current distributions) of an asymptotically vanishing electromagnetic
radiation field can be determined or at least studied from the asymptotic
field itself, e.g. the total charge and both the electric and magnetic
multipole moments. In GR one tries to do a similar thing. For example we
have the Bondi energy-momentum defined from certain asymptotic Weyl tensor
components and many attempts to define multipole moments and angular
momentum from other Weyl tensor components and the asymptotic shear.

Our contention is that, aside from the energy-momentum four-vector, many or
most of the asymptotically obtained physical quantities are hidden in the
UCF. And even the properties and description of the energy-momentum
four-vector is greatly clarified by the UCF. In order to extract this
information from the UCF , using its assumed regularity, we expand it in
spherical harmonics, writing it as 
\[
u=X(\tau ,\zeta ,\widetilde{\zeta })=\Sigma \xi ^{lm}(\tau )Y_{lm}(\zeta ,%
\widetilde{\zeta }), 
\]
or, using the spherical tensor harmonics\cite{spins} as, 
\begin{equation}
u=X(\tau ,\zeta ,\widetilde{\zeta })=\frac{1}{\sqrt{2}}\xi ^{0}(\tau
)Y_{0}^{0}(\zeta ,\widetilde{\zeta })-\frac{1}{2}\xi ^{i}(\tau
)Y_{1i}^{0}(\zeta ,\widetilde{\zeta })+\xi ^{ij}(\tau )Y_{1ij}^{0}(\zeta ,%
\widetilde{\zeta })+...,  \label{expansion}
\end{equation}
where the components $\xi ^{lm}(\tau )$ or the ($\xi ^{0}(\tau ),\xi
^{i}(\tau ),\xi ^{ij}(\tau ),.....$) are, in general, complex functions of a
complex time $\tau ,$ and are to be given physical meaning. Eventually, the $%
\tau $ must be replaced, via Eq.(\ref{T}), by the real $u$. The four
components of ($\xi ^{0},\xi ^{1m}$)$\Leftrightarrow $($\xi ^{0},\xi ^{i})$
are to be identified with a complex position vector, the real part with the
center of mass, and the imaginary part with spin-angular momentum/unit mass.
The complex $\xi ^{2m}\Leftrightarrow \xi ^{ij}$ are associated with mass
and spin-quadrupole moments while the higher harmonic terms are related to
the higher moments, the real parts to the mass moments and the imaginary
parts to the spin-moments.

In addition to its role as describing physical quantities, the UCF has a
pretty geometric/group theoretical meaning.

One knows from Bondi's and Penrose's early work concerning the asymptotic
symmetries\cite{RP,Sachs,RP2,NT} that the freedom in the choice of the Bondi
coordinate system is what is known as the BMS group. In addition to the
rotation subgroup it contains the supertranslation subgroup acting on $\frak{%
\ I}^{+}$ and given by 
\[
u=u^{\prime }+\alpha (\zeta ,\overline{\zeta }), 
\]
which can be extended into the complex by 
\[
u=u^{\prime }+\alpha (\zeta ,\widetilde{\zeta }). 
\]
The function $\alpha (\zeta ,\widetilde{\zeta }),$ which can be expanded in
spherical harmonics, contains the (complex) Poincar\'{e} translation
subgroup defined by the expansion in only the $l=0$ and $l=1$ harmonics, 
\begin{equation}
\alpha (\zeta ,\widetilde{\zeta })=\xi ^{0}Y_{0}^{0}(\zeta ,\widetilde{\zeta 
})+\xi ^{i}Y_{1i}^{0}(\zeta ,\widetilde{\zeta }).  \label{Poincare}
\end{equation}
If we choose the starting value $u^{\prime }=0$ and consider a one parameter
family of supertranslations, $\alpha (\zeta ,\widetilde{\zeta })\Rightarrow
\alpha (\tau ,\zeta ,\widetilde{\zeta }),$ we can consider the UCF as a one
complex parameter family of supertranslations and the $l=0$ and $l=1$ part
as a one-parameter family of complex Poincar\'{e} transformations.

In other words, the complex four-vector world-line $\xi ^{a}(\tau )=(\xi
^{0}(\tau ),\xi ^{i}(\tau ))$ can be interpreted as a world-line in the
(complex) Poincar\'{e} translation subgroup. We can chose $\tau $ so that
the velocity vector of the world-line has a Lorentzian norm of one. \{An
alternative interpretation is to take the world-line in H-space and use the
H-space norm.\}

The remainder of this work is organized as follows: In Sec. III, the
application of the UCF to Minkowski space is described and the two different
flat space forms of the metric, those associated with $L(u,\zeta ,\overline{
\zeta })$ and with $V(\tau ,\zeta ,\overline{\zeta }),$ are worked out. In
Sec. IV we give a brief outline how the UCF is used in general
asymptotically flat space-times while in Secs. V-VII we apply these idea to
the algebraically special type II metrics. Sec. VIII is devoted to a
discussion.

\section{The UCF and Flat Space}

Since many of the ideas developed here are unfamiliar to large portions of
the relativity community we believed it would be of value to first describe
them in the simpler context of flat space-time where the functions $L$ and $%
V $ have made frequent earlier appearances\cite{NT,RTM} and have relatively
simple meaning.

The basic idea here is to first describe arbitrary shear-free null geodesic
congruences in flat space and then to specialize them to the case of regular
congruences, i.e., congruences that are completely transverse to $\frak{I}
^{+.}.$ In general these congruences have twist and the basic variable of
choice is the $L$. Twist-free congruences are more easily treated with the
variable $V$. The main distinction between the flat space and the
asymptotically flat space properties of $L$ and $V$ is that in the spherical
harmonic expansions only the lowest harmonics, i.e., $l=0$ and $1,$ appear
in the flat space while all harmonics can appear for the curved space.

We begin by giving an analytic description of any arbitrary null geodesic
congruence, (with $l^{a}$ as tangent vector), in Minkowski space\cite{Ntwist}
. Using $x^{a}$ as standard Minkowski-space coordinates, $L(u,\zeta ,%
\overline{\zeta })$ an arbitrary complex function of the parameters ($%
u,\zeta ,\overline{\zeta }$) which label individual members of the null
geodesic congruence and $r$ as the affine parameter along each null
geodesic, we find that 
\begin{eqnarray}
x^{a} &=&u_{r}t^{a}-L\overline{m}^{a}-\overline{L}m^{a}+(r-r_{0})l^{a},
\label{ngc} \\
x^{a} &=&u_{r}\frac{(l^{a}+n^{a})}{\sqrt{2}}-L\overline{m}^{a}-\overline{L}%
m^{a}+(r-r_{0})l^{a},  \label{ngc'} \\
x^{a} &=&u(l^{a}+n^{a})-L\overline{m}^{a}-\overline{L}m^{a}+(r-r_{0})l^{a},
\label{ngc''} \\
u_{r} &=&\sqrt{2}u,
\end{eqnarray}
with the null tetrad 
\begin{eqnarray}
l^{a} &=&\frac{\sqrt{2}}{2P}(1+\zeta \overline{\zeta },\text{ }\zeta +%
\overline{\zeta },\text{ }-i(\zeta -\overline{\zeta }),-1+\zeta \overline{
\zeta });  \label{tetrad} \\
m^{a} &=&\text{\dh }l^{a}=\frac{\sqrt{2}}{2P}(0,1-\overline{\zeta }^{2},-i(1+%
\overline{\zeta }^{2}),\text{ }2\overline{\zeta }),  \nonumber \\
\overline{m}^{a} &=&\overline{\text{\dh }}l^{a}=\frac{\sqrt{2}}{2P}
(0,1-\zeta ^{2},\text{ }i(1+\zeta ^{2}),2\zeta ),  \nonumber \\
n^{a} &=&\frac{\sqrt{2}}{2P}(1,-\zeta +\overline{\zeta },\text{ }i(\zeta -%
\overline{\zeta }),1-\zeta \overline{\zeta }),  \nonumber \\
t^{a} &=&\frac{l^{a}+n^{a}}{\sqrt{2}},  \label{t} \\
P &=&1+\zeta \overline{\zeta }.  \label{P}
\end{eqnarray}

For the simplification of the complex divergence of the congruence the
function $r_{0}$ is most often chosen as 
\begin{equation}
r_{0}=-\frac{1}{2}\{\text{\dh }\overline{L}+L\overline{L}^{{\large \cdot }}+%
\overline{\text{\dh }}L+\overline{L}L^{{\large \cdot }}\}.  \label{r_0}
\end{equation}

Our main interest lies in the class of regular null geodesic congruences
with a vanishing shear. This condition is achieved by first imposing the
shear-free condition on $L,$ i.e., the differential condition\cite{A}

\begin{equation}
\text{\dh }L+LL^{{\large \cdot }}=0,  \label{Shearfree}
\end{equation}
and then requiring solutions to be regular functions (expandable in
spherical harmonics) on the ($\zeta ,\overline{\zeta }$) sphere. To solve
this equation we introduce the (complex) auxiliary function, \{a CR
function\} 
\begin{equation}
\tau =T(u,\zeta ,\overline{\zeta }),
\end{equation}
by

\begin{equation}
\text{\dh }T+LT^{\,{\large \cdot }}=0.  \label{ethTarXivmod}
\end{equation}

By using the inverse function to $\tau =T(u,\zeta ,\overline{\zeta }),$
i.e., the UCF

\begin{equation}
u=X(\tau ,\zeta ,\overline{\zeta }),  \label{X*}
\end{equation}
and implicit differentiation, $L$ can be written as 
\begin{equation}
L=\text{\dh }_{(\tau )}X(\tau ,\zeta ,\overline{\zeta }),  \label{L*}
\end{equation}
and Eq.(\ref{Shearfree}) becomes

\begin{equation}
\text{\dh }_{(\tau )}^{2}X=0.  \label{eth^2X}
\end{equation}

The general regular solution, $L(u,\zeta ,\overline{\zeta }),$ to Eq.(\ref
{Shearfree}) is then given in the parametric form 
\begin{eqnarray}
L &=&\xi ^{a}(\tau )m_{a}(\zeta ,\overline{\zeta }),  \label{L**} \\
u &=&\xi ^{a}(\tau )l_{a}(\zeta ,\overline{\zeta }),  \label{X**}
\end{eqnarray}
where $\xi ^{a}(\tau )$ are four arbitrary \textit{complex} functions of $%
\tau .$ Since Eq.(\ref{T}) is invariant under $T^{*}=F(T),$ we can take $%
\tau $ so that 
\[
v^{a}(\tau )=\xi ^{a}(\tau )^{\prime }=\partial _{\tau }\xi ^{a}, 
\]
is a unit$\ $Lorentzian four-vector, $v^{a}v_{a}=1$. Using the spin-s tensor
notation\cite{RTM,spins}

\begin{eqnarray*}
l_{a} &=&(\frac{1}{\sqrt{2}},-\frac{1}{2}Y_{1i}^{0}), \\
m_{a} &=&(0,Y_{1i}^{1}),
\end{eqnarray*}
Eqs.(\ref{L**}) and (\ref{X**}) become 
\begin{eqnarray}
L &=&\xi ^{i}(\tau )Y_{1i}^{1},  \label{L***} \\
u &=&\frac{\xi ^{0}(\tau )}{\sqrt{2}}-\frac{1}{2}\xi ^{i}(\tau )Y_{1i}^{0}.
\label{X***}
\end{eqnarray}

\begin{remark}
We comment again that $L$ is a complex function that describes a
stereographic `angle' field on real $\frak{I}^{+},$ i.e., a direction field
that determines a null geodesic congruence. It is generated by the UCF that
is determined by an arbitrary complex world-line in complex Minkowski space 
\[
z^{a}=\xi ^{a}(\tau ). 
\]
When $L$ from Eqs.(\ref{L***}) and (\ref{X***}) is used in Eq.(\ref{ngc''}),
we obtain the explicit form for any regular shear-free null geodesic
congruence.
\end{remark}

Later, in more general situations, we will use this form, Eqs.(\ref{L***})
and (\ref{X***}), for the first two harmonics. Curvatures come from the
higher harmonics.

An alternative meaning of Eq.(\ref{ngc''}) is that it defines a coordinate
transformation from the natural coordinates of the congruence ($u,r,\zeta 
\overline{,\zeta }$) to the Minkowski coordinates, $x^{a}$. One can in a
straightforward way find the flat metric in these coordinates.

From Eq.(\ref{ngc''}), calculating the differentials we find 
\begin{eqnarray}
dx^{a} &=&(1-r_{0}^{{\large \cdot }})l^{a}du+n^{a}du+l^{a}dr-\overline{L}^{%
{\large \cdot }}m^{a}du-L^{{\large \cdot }}\overline{m}^{a}du  \label{dx} \\
&&+P^{-1}[(r-r_{0}-\overline{\text{\dh }}L)\overline{m}^{a}-\overline{\text{
\dh }}\overline{L}m^{a}-\overline{L}n^{a}+(\overline{L}-\overline{\text{\dh }%
}r_{0})l^{a}]d\overline{\zeta }  \nonumber \\
&&+P^{-1}[(r-r_{0}-\text{\dh }\overline{L})m^{a}-\text{\dh }L\overline{m}
^{a}-Ln^{a}+(L-\text{\dh }r_{0})l^{a}]d\zeta .  \nonumber
\end{eqnarray}
Since the set ($l=l_{a}dx^{a},n=n_{a}dx^{a},m=m_{a}dx^{a},\overline{m}=%
\overline{m}_{a}dx^{a}$) define a null one-form system the metric can be
written as 
\begin{equation}
\eta =\eta _{ab}dx^{a}dx^{b}=2(ln-m\overline{m}).  \label{metric}
\end{equation}
Any other set of one-forms obtained from the first set by a Lorentz
transformation yields the same metric. Choosing a new null tetrad via the
null rotation 
\begin{eqnarray*}
l^{*} &=&l, \\
m^{*} &=&m-L^{{\large \cdot }}l, \\
\overline{m}^{*} &=&\overline{m}-\overline{L}^{{\large \cdot }}l, \\
n^{*} &=&n-\overline{L}^{{\large \cdot }}m-L^{{\large \cdot }}\overline{m}+%
\overline{L}^{{\large \cdot }}L^{{\large \cdot }}l,
\end{eqnarray*}
we obtain, by a direct (lengthy) calculation, using Eq.(\ref{Shearfree}),
the simplest form of the tetrad, namely 
\begin{eqnarray}
l^{*} &=&dx^{a}l_{a}=du-P^{-1}(Ld\zeta +\overline{L}d\overline{\zeta }),
\label{tetrad*} \\
m^{*} &=&-P^{-1}[r-i\Sigma ]d\overline{\zeta }=P^{-1}\overline{\rho }^{-1}d%
\overline{\zeta },  \nonumber \\
\overline{m}^{*} &=&-P^{-1}[r+i\Sigma ]d\zeta =P^{-1}\rho ^{-1}d\zeta , 
\nonumber \\
n^{*} &=&dr+P^{-1}\{K+L^{{\large \cdot }}\overline{\rho }^{-1})d\zeta +(%
\overline{K}+\overline{L}^{{\large \cdot }}\rho ^{-1})d\overline{\zeta }%
\}+(1+\frac{1}{2}\{\text{\dh }\overline{L}^{{\large \cdot }}+\overline{\text{
\dh }}L^{{\large \cdot }}+L\overline{L}^{{\large \cdot \cdot }}+\overline{L}%
L^{{\large \cdot \cdot }}\})l,  \nonumber
\end{eqnarray}
where 
\begin{eqnarray}
\rho &=&-\frac{1}{r+i\Sigma },  \label{rho} \\
i\Sigma &\equiv &-r_{0}-\overline{\text{\dh }}L-L^{{\large \cdot }}\overline{%
L}\equiv \frac{1}{2}\{\text{\dh }\overline{L}+L\overline{L}^{{\large \cdot }%
}-\overline{\text{\dh }}L-\overline{L}L^{{\large \cdot }}\},  \label{sigma}
\\
K &=&L+L\text{\dh }\overline{L}^{{\large \cdot }}+\frac{1}{2}\{\text{\dh }
^{2}\overline{L}+L^{2}\overline{L}^{{\large \cdot \cdot }}+L^{{\large \cdot }
}[\text{\dh }\overline{L}-\overline{\text{\dh }}L+L\overline{L}^{{\large %
\cdot }}-\overline{L}L^{{\large \cdot }}]\},  \label{K}
\end{eqnarray}
with $\rho $ being the complex divergence of the congruence and $\Sigma $
its twist. The metric is then given by. 
\begin{equation}
\eta =2(l^{*}n^{*}-m^{*}\overline{m}^{*}).  \label{metric*}
\end{equation}

This form of the metric is very close to that the Type II twisting metrics%
\cite{NT}.

If the world-line $\xi ^{a}(\tau )$ was described by a \textit{real function
of a real} $\tau ,$ though Eqs.(\ref{tetrad*}) and (\ref{metric*}) could
still be used \{with now $\Sigma =0$\}, there however is a simpler set of
coordinates, namely ($\tau ,r,\zeta ,\overline{\zeta }$), associated with
the non-twisting congruence, that can be used\cite{RTM}. They are given by 
\begin{equation}
x^{a}=\xi ^{a}(\tau )+\frac{r}{V}l^{a}(\zeta ,\overline{\zeta }),
\label{ngc^}
\end{equation}
where 
\begin{eqnarray}
u &=&\xi ^{a}{}(\tau )l_{a}=X(\tau ,\zeta ,\overline{\zeta }),  \label{X^} \\
V &\equiv &X^{\prime }=\xi ^{a}{}^{\prime }l_{a}=v^{a}{}l_{a}=\frac{v^{0}}{%
\sqrt{2}}-\frac{1}{2}v^{i}Y_{1i}^{0}.  \nonumber
\end{eqnarray}
By a computation similar to that with the complex curve we obtain the flat
metric in $(\tau ,r,\zeta ,\overline{\zeta })$ coordinates, 
\begin{equation}
ds^{2}=\eta _{ab}dx^{a}dx^{b}=(1-2\frac{V^{\prime }}{V}r)d\tau ^{2}+2d\tau
dr-r^{2}\frac{2d\zeta d\overline{\zeta }}{V^{2}P^{2}},  \label{flat metric}
\end{equation}
which is very similar in form to the type II (non-twisting)
Robinson-Trautman metrics\cite{RT,RTM}..

As we pointed out, in flat space the UCF is \textit{not unique}. It is
determined, in general, by an arbitrary complex world-line in complex
Minkowski space. However in the presence of an asymptotically vanishing
Maxwell field (with non-vanishing total charge) one can construct, from the
Maxwell field itself, a unique complex world-line (the complex center of
charge)\cite{KN} and from it a unique UCF can be found. For example, the
Lienard-Wiechert charged world-line defines\cite{KN,gyro} the real $\xi
^{a}{}(\tau )$.

In the case of asymptotically flat vacuum space-times, properties of the
Weyl tensor are used to define the world-line\cite{KNO,KNO2,FtPrints,KN} and
the unique UCF. In the case of asymptotically flat Einstein-Maxwell
space-times there are in general two different complex world-lines,
determined respectively by the properties of the Weyl tensor and the Maxwell
field. As we said earlier, we here consider only the case where the two
lines coincide.

\section{Asymptotically Flat Space-Times}

In this section we outline - leaving out many of the details - how the UCF
is determined in asymptotically flat space-times.

For asymptotically flat space-times, the basic radiation data is the Bondi
asymptotic shear, $\sigma (u,\zeta ,\overline{\zeta })$, a complex function
given on $\frak{I}^{+}$ which, as we mentioned earlier, we assume can be
analytically extended a short way into $\frak{I}_{C}^{+}$, and hence written
as $\sigma (u,\zeta ,\widetilde{\zeta })$

The $\sigma (u,\zeta ,\widetilde{\zeta })$ describes the shear of the null
geodesic congruence that are the null generators of the Bondi null surfaces
given by $u=constant$ in the neighborhood of $\frak{I}^{+}$. Any new \textit{%
asymptotic} null geodesic congruence can be obtained by a Lorentz
transformation (specifically a null rotation around the null generators of $%
\frak{I}_{C}^{+})$ at $\frak{I}^{+}$. If $l^{a}$ are the tangent vectors to
the Bondi congruence the new tangent vectors, $l^{*\,a},$ in the
neighborhood of $\frak{I}^{+}\ $are given by\cite{FtPrints,KNO,KNO2} 
\begin{eqnarray}
l^{*a} &=&l_{B}^{a}-\frac{L}{r}\overline{m}_{B}^{a}+\frac{\overline{L}}{r}%
m_{B}^{a}+O(r^{-2}),  \label{NullRot} \\
m^{*a} &=&m_{B}^{a}-\frac{L}{r}n_{B}^{a}+O(r^{-2}),  \nonumber \\
n^{*a} &=&n_{B}^{a},  \nonumber
\end{eqnarray}
with $L(u,\zeta ,\overline{\zeta })$, at this point, an arbitrary complex
function on $\frak{I}^{+}.$ The asymptotic shear of the new congruence is
given by\cite{A} 
\begin{equation}
\sigma ^{*}=\sigma -\text{\dh }_{(u)}L-LL^{\,{\large \cdot }}.
\label{shear*}
\end{equation}
By requiring that the new congruence have vanishing asymptotically shear, we
see that $L(u,\zeta ,\overline{\zeta })$ must satisfy the differential
equation, Eq.(\ref{ethL*}),

\begin{equation}
\text{\dh }_{(u)}L(u,\zeta ,\widetilde{\zeta })+LL^{{\large \cdot }}=\sigma
(u,\zeta ,\widetilde{\zeta }),  \label{ethL**}
\end{equation}
where now we are considering the extension into the complex. The technique
for studying properties of solutions of Eq.(\ref{ethL**}) is quite similar
to what was done in the Minkowski space case, Eq.(\ref{Shearfree}).

Again we introduce the new complex auxiliary function 
\begin{equation}
\tau =T(u,\zeta ,\widetilde{\zeta }),  \label{T**}
\end{equation}
by 
\begin{equation}
L=-\frac{\text{\dh }_{(u)}T}{T^{\,{\large \cdot }}},  \label{eth^}
\end{equation}
and its inverse 
\begin{equation}
u=X(\tau ,\zeta ,\widetilde{\zeta }),  \label{UCF}
\end{equation}
which eventually leads to the UCF. \{Note that we have the useful gauge
freedom $\tau ^{*}=T^{*}(u,\zeta ,\widetilde{\zeta })=F(T(u,\zeta ,%
\widetilde{\zeta }))$ in the determination of $T(u,\zeta ,\widetilde{\zeta }
).\}$ By implicit differentiation of Eq.(\ref{UCF}), i.e., 
\begin{eqnarray*}
1 &=&X^{\prime }T^{\,{\large \cdot }}, \\
0 &=&\text{\dh }_{(\tau )}X+X^{\prime }\text{\dh }_{(u)}T,
\end{eqnarray*}
we find the implicit form of $L(u,\zeta ,\widetilde{\zeta })$%
\begin{eqnarray}
L &=&\text{\dh }_{(\tau )}X(\tau ,\zeta ,\widetilde{\zeta }),  \label{ethX}
\\
u &=&X(\tau ,\zeta ,\widetilde{\zeta }).  \label{UCF*}
\end{eqnarray}
Continuing with the change of variable, Eq.(\ref{ethL**}) becomes 
\begin{equation}
\text{\dh }_{(\tau )}^{2}X=\sigma (X,\zeta ,\widetilde{\zeta }),
\label{goodcut}
\end{equation}
a well studied equation, frequently referred to as the `good-cut' equation%
\cite{KNT,HNPT}. Its solution, which depend on four complex parameters, $
z^{a}$ (defining H-space) is written as 
\[
X=\widehat{X}(z^{a},\zeta ,\widetilde{\zeta }). 
\]
Now by taking an arbitrary world-line in the parameter space, $z^{a}=\xi
^{a}(\tau ),$ we have the parametric form of the solution to Eq.(\ref{ethL**}
), 
\begin{eqnarray*}
u &=&X(\tau ,\zeta ,\widetilde{\zeta })=\widehat{X}(\xi ^{a}(\tau ),\zeta ,%
\widetilde{\zeta }), \\
L &=&\text{\dh }_{(\tau )}X(\tau ,\zeta ,\widetilde{\zeta }).
\end{eqnarray*}
It is important to notice \{see Eq.(\ref{X***})\} that in the harmonic
expansion of $X(\tau ,\zeta ,\widetilde{\zeta })$ we can take the first four
harmonic coefficients, $l=(0,1)$, as the $\xi ^{a}(\tau ),$ i.e., write 
\begin{eqnarray}
u &=&\xi ^{a}(\tau )l_{a}+harmonics(l\geq 2),  \label{expansion**} \\
u &=&\frac{\xi ^{0}}{\sqrt{2}}(\tau )-\frac{1}{2}\xi ^{i}(\tau
)Y_{1i}^{0}+harmonics(l\geq 2),  \nonumber
\end{eqnarray}
and that 
\begin{equation}
V=X^{\prime }=\frac{v^{0}}{\sqrt{2}}(\tau )-\frac{1}{2}v^{i}(\tau
)Y_{1i}^{0}+harmonics(l\geq 2).  \label{Vexpansion}
\end{equation}

We see that in both the flat-space case and for asymptotically flat
space-times the freedom in choosing a regular shear-free or asymptotically
shear- free null geodesic congruence lies in the arbitrary choice of a
complex world-line in the parameter space, i.e., complex Minkowski space or
H-space - or via the alternative interpretation, the space of complex
Poincar\'{e} translations.

The issue is: \textit{can one uniquely determine this world-line}?

\textit{In Minkowski space the answer is no, except when there is an
asymptotically vanishing charged Maxwell field}. Using the null rotation,
Eq.(\ref{NullRot}), and requiring the vanishing of the coefficient of the $%
l=1$ harmonic of the Maxwell tetrad field, $\phi _{0}^{*0},$ (the complex
dipole term), one determines a unique world-line, the `complex center of
charge'\cite{gyro}.

\textit{For asymptotically flat spaces}, with a similar null rotation, and
then requiring the vanishing of the $l=1$ harmonic of the Weyl tensor
component, $\Psi _{1}^{0},$ i.e., the complex mass dipole terms, \textit{the
world-line } 
\[
z^{a}=\xi ^{a}(\tau ), 
\]
\textit{is again determined}\cite{KNO,KNO2}.As the details for the
determination of the unique complex curve have been described earlier\cite
{KNO,KNO2} and a further discussion would take us far afield, we will not
pursue this issue here.

\begin{remark}
We note that the information in the Bondi shear begins with the $l=2$, i.e.,
quadrupole, harmonic while the UCF, in addition to its higher harmonics, $%
l\eqslantgtr $ 2, also contains the information in the $l=0,1$ harmonics
obtained from the unique world-line. In other words while the UCF uniquely
determines the Bondi shear, Eq.(\ref{goodcut}), the Bondi shear only
determines the higher harmonics of the UCF. In essence the new structures we
have been describing here are just these $l=0,1$ harmonics of the UCF or
equivalently, the complex world-line. Later we will see the physical content
contained in this new information.
\end{remark}

As an aside we note that in the very special case, when the unique
asymptotically shear free congruence is also twist-free, the UCF, $u=X(\tau
,\zeta ,\overline{\zeta }),$ will be a real function of a real $\tau .$ It
could then be interpreted as simply the coordinate transformation between
Bondi time and NU time.

\begin{remark}
To end this section, we describe the natural relationship our UCF has with
the elegant mathematical idea of a CR structure\cite{CR,CRNew,Tr,RP3,CR1,CR2}
. The UCF defines or induces a CR structure on the real $\frak{I}^{+}$ in
the following fashion: from the UCF we obtain $L(u,\zeta ,\bar{\zeta}),$ as
above, but now restrict $(u,\zeta ,\bar{\zeta})$ to the real $\frak{I}^{+}.$
Then define the CR structure by (the equivalence class of) the three
one-forms, one real and the complex conjugate pair: 
\begin{eqnarray}
l^{*}=du-\frac{L}{1+\zeta \bar{\zeta}}\mathrm{d}\zeta -\frac{\bar{L}}{%
1+\zeta \bar{\zeta}}\mathrm{d}\bar{\zeta},~~~  \label{one-forms} \\
~m^{*}=\frac{\mathrm{d}\overline{{\zeta }}}{1+\zeta \bar{\zeta}},\qquad ~%
\overline{m}^{*}=\frac{\mathrm{d}{\zeta }}{1+\zeta \bar{\zeta}}.  \nonumber
\end{eqnarray}

Writing $\zeta =x+iy,$ the functions $\tau =T(u,x+iy,x-iy)$ and $\widetilde{%
\zeta }=x-iy$ are the CR functions defining the embedding of $\frak{I}^{+}$
into \textbf{C}$^{2}=\{\tau ,\widetilde{\zeta }\},$ both satisfying the CR
equation, i.e., \dh $_{(u)}T+LT^{\cdot }=0,$ etc., which was originally our
defining equation for $T.$ It is natural since the complex conjugate pair ($%
~m^{*},~\overline{m}^{*}$) are the dual covectors to the $u=constant$
surfaces of $\frak{I}^{+}$ and $l^{*}$ is the real one-form determining the
null direction of the new asymptotically shear-free null geodesic congruence.
\end{remark}

Though we will not make further direct use of the CR structure we note that
it is the CR structure that lies at the basis of much of the present work.
The function $\tau =T(u,\zeta ,\bar{\zeta})$ essentially defines the CR
structure and its inverse, $u=X(\tau ,\zeta ,\bar{\zeta}),$ yields the
physical content of the work.

\section{Type II Metrics and the UCF}

\subsection{Non-Twisting:The Robinson-Trautman Case}

Though our issue here is the analysis of the type II twisting metrics - with
and without a Maxwell field - we nevertheless want to begin with a brief
discussion of the non-twisting case, i.e., the Robinson-Trautman metrics.

Since Robinson-Trautman (RT) metrics have been extensively studied and
written about\cite{RT,Piotr,RTM}, we will confine ourselves to the
discussion of their relationship with the UCF.

Rather than beginning with the differential equations for the determination
of the Robinson-Trautman (RT) metrics, we reverse the proceedings. We will
assume that we have the general solution to the RT equations and that
furthermore we explicitly know the unique UCF, 
\begin{equation}
u=X_{RT}(\tau ,\zeta ,\bar{\zeta}),  \label{X_RT}
\end{equation}
which in this case is real with real $\tau .$ The $(\tau ,\zeta ,\bar{\zeta}
) $ are the RT/NU coordinates on $\frak{I}^{+},$ with the UCF, Eq.(\ref{X_RT}%
), having the added meaning of being the coordinate transformation to Bondi
coordinates. The function 
\[
V_{RT}=X_{RT}^{\prime }(\tau ,\zeta ,\bar{\zeta}), 
\]
then is the solution to the RT equations 
\begin{eqnarray}
\text{\dh }_{(\tau )}\chi &=&0,  \label{RT} \\
\chi ^{\prime }-3\chi \frac{V_{RT}^{\prime }}{V_{RT}}-V_{RT}^{3}\{\text{\dh }
_{(\tau )}^{2}\overline{\text{\dh }}_{(\tau )}^{2}V_{RT}-V_{RT}^{-1}%
\overline{\text{\dh }}_{(\tau )}^{2}V_{RT}\cdot \text{\dh }_{(\tau
)}^{2}V_{RT}\} &=&0,  \label{chi}
\end{eqnarray}
\begin{equation}
\chi \equiv -V_{RT}^{3}\psi _{2}^{*0}.  \label{chi*}
\end{equation}
where $\psi _{2}^{*0}$ is a Weyl tensor component. From Eq.(\ref{RT}) we see
that $\chi $ depends only on $\tau .$ Using the freedom of time change, $%
\tau =f(\tau ^{*}),$ $\chi $ could be made constant. However this freedom
was already used to make $v^{a}(\tau )=\xi ^{a\,\prime }$ into a unit vector
so that $\chi $ remains a function of $\tau .$ The time evolution, Eq.\ref
{chi}), is in fact equivalent to the Bondi energy-momentum loss equation.
When Eq.(\ref{chi}) is transformed to Bondi coordinates via Eq.(\ref{X_RT})
it becomes 
\[
\Psi ^{{\large \cdot }}=\sigma ^{{\large \cdot }}(\overline{\sigma })^{%
{\large \cdot }\,}, 
\]
with $\Psi ,$ the mass aspect given by 
\begin{equation}
\Psi =\psi _{2}^{*0}+2L\text{\dh }(\overline{\sigma })^{{\large \cdot }
}+L^{2}(\overline{\sigma })^{{\large \cdot \cdot }}+\text{\dh }^{2}\overline{
\sigma }+\sigma (\overline{\sigma })^{{\large \cdot }\,}=\overline{\Psi }.
\label{MassAspect}
\end{equation}
The $L$ and $\sigma $ are directly calculated from the UCF as described
earlier.

Though it is virtually impossible to exactly perform the coordinate
transformation to Bondi coordinates, one can do so within an approximation
scheme. Using a second order expansion, essentially in $v/c$, one finds from
the mass aspect that the Bondi mass and momentum can be partially expressed
in terms of the world-line and take on the surprising familiar kinematically
form 
\begin{eqnarray}
\text{ }M &=&M_{0}[1+\frac{v^{i2}}{2c^{2}}+\frac{v^{i\prime }\xi ^{i}}{c^{2}}
]+\text{quadrupole terms,}  \label{M&P} \\
P^{i} &=&M_{0}v^{i}+\text{quadrupole terms,}  \nonumber
\end{eqnarray}
where we see the approximation to the relativistic mass 
\[
M=\frac{M_{0}}{\sqrt{1-\frac{v^{i2}}{c^{2}}}}\cong M_{0}[1+\frac{v^{i2}}{
2c^{2}}], 
\]
with the strange term $M_{0}\frac{v^{i\prime }\xi ^{i}}{c^{2}}$ appearing to
be the ``work done'' moving the ``particle'' to the ``position ``$\xi ^{i}".$

The Bondi mass-momentum loss equations, to second order, become the
equations of motion for the world-line. 
\begin{eqnarray}
M^{\prime } &=&-\frac{1}{5c^{7}}GQ^{ij\prime \prime \prime }Q^{ij\prime
\prime \prime },  \label{M&P loss} \\
Mv^{i\,\prime } &=&v^{i}\frac{G}{5c^{7}}Q^{jk\prime \prime \prime
}Q^{jk\prime \prime \prime },  \label{motion}
\end{eqnarray}
where $Q$ is defined from the $l=2$ coefficient in the harmonic expansion of 
$X,$ Eq.(\ref{expansion}) by

\begin{equation}
\xi ^{ij}=\frac{G}{\sqrt{2}12c^{4}}Q^{ij\prime \prime }.  \label{Q}
\end{equation}

These results, when generalized to the Robinson-Trautman-Maxwell equations%
\cite{RTM}, contain further surprising results, e.g., Eq.(\ref{motion})
contains, among other terms, the classical radiation reaction force given by 
\[
\frac{2q^{2}}{3c^{3}}\xi ^{i\prime \prime \prime }. 
\]

See also Eq.(\ref{EqsOfMotionII}) below.

\subsection{Type II Twisting Metrics}

\subsubsection{Solving the field equations}

Our primary objective in this work is to analyze and look for the physical
meaning of the different geometric variables that arise in the vacuum and
Einstein-Maxwell type II twisting metrics. In particular we want to see how
the UCF enters into the discussion. The relevant final differential
equations to determine the metrics have been known and studied for many years%
\cite{Lind,Talbot,NT,RTM}. Though a great deal of mathematical insight,
specifically the discovery and understanding of the associated CR structure,
was developed, aside from the Kerr metric and the charged Kerr metric,
general physical understanding of these metrics seems to be missing. Here,
with the help of the UCF, this lack of understanding will be addressed.

Acknowledging the non-linearities and the difficulties in finding exact
results, we approach the problem of physical interpretation using, in
addition to the UCF, the two approximations: (1) In the spherical harmonic
expansions only terms with $l=(0,1,2)$ harmonics will be retained and (2)
calculations will be done to second order in $v/c$.

The Type II coupled differential equations for the independent variables $%
(L,\phi _{1}^{*0},\phi _{2}^{*0},\psi _{2}^{*0})$ in terms of $(u,\zeta ,%
\overline{\zeta })$ \cite{NT,KNO,KNO2,KN} 
\begin{eqnarray}
\text{\dh }\phi _{1}^{*0}+2L^{{\large \cdot }}\phi _{1}^{*0}+L(\phi
_{1}^{*0})^{{\large \cdot }} &=&0,  \label{m1} \\
(\phi _{1}^{*0})^{{\large \cdot }}+\text{\dh }\phi _{2}^{*0}+(L\phi
_{2}^{*0})^{{\large \cdot }} &=&0,  \label{m2}
\end{eqnarray}

\begin{eqnarray}
0 &=&\text{\dh }\psi _{2}^{*0}+L(\psi _{2}^{*0\,})^{{\large \cdot }}+3L^{%
{\large \cdot }}\psi _{2}^{*0}-2k\phi _{1}^{*0}\overline{\phi }_{2}^{*0},
\label{gr1} \\
\Psi ^{{\large \cdot }} &=&\sigma ^{{\large \cdot }}(\overline{\sigma })^{%
{\large \cdot }}+k\phi _{2}^{*0}\overline{\phi }_{2}^{*0},  \label{gr2} \\
\Psi &=&\overline{\Psi }=\psi _{2}^{*0}+2L\text{\dh }(\overline{\sigma })^{%
{\large \cdot }}+L^{2}(\overline{\sigma })^{{\large \cdot \cdot }}+\text{\dh 
}^{2}\overline{\sigma }+\sigma (\overline{\sigma })^{{\large \cdot }},
\label{gr3}
\end{eqnarray}
with 
\begin{equation}
\sigma \equiv \text{\dh }L+LL^{{\large \cdot }}.  \label{SIGMA}
\end{equation}
Eq.(\ref{gr3}) defines the mass aspect with the requirement that it be real
while Eq.(\ref{SIGMA}) defines the Bondi shear $\sigma $ from the $L$. Eq.(%
\ref{gr2}) is the evolution equation for the Bondi mass aspect with the $%
l=(0,1)$ terms determining the energy and momentum loss. Basically these
equations have the following structure: from the two Maxwell equations, Eqs.(%
\ref{m1}) and (\ref{m2}), both $\phi _{1}^{*0}$ and $\phi _{2}^{*0}$ will be
determined in terms of the $L$ and a constant of integration, $q$, the
charge. Eq.(\ref{gr1}) determines $\psi _{2}^{*0}$ (or $\Psi $) in terms of $%
L$ and a single function $\Upsilon (u$)$.$ The function $L(u,\zeta ,%
\overline{\zeta })$ could then be considered as our basic variable which
would be determined by Eq.(\ref{gr2}). We however will consider something
different. Our point of view is that $L$ is a derived function, coming from
the UCF as was described earlier. Since it is the information encoded in the
UCF that we want to use, we must reexpress the differential equations in
terms of the UCF. Our first task will be to transform three of the field
equations, i.e., Eqs.(\ref{m1}), (\ref{m2}) and \ref{gr1}), to a new very
useful form.

Assuming that we know the UCF, $u=X(\tau ,\zeta ,\widetilde{\zeta }),$ then
and its inverse, $\tau =T(u,\zeta ,\widetilde{\zeta })$ and $T^{\cdot
}=\partial _{u}T$ contain the information from $X.$ We define $\Upsilon
(u,\zeta ,\widetilde{\zeta }),\Phi _{1}(u,\zeta ,\widetilde{\zeta }),\Phi
_{2}(u,\zeta ,\widetilde{\zeta })$ by 
\begin{eqnarray}
\psi _{2}^{*0} &=&(T^{\,{\large \cdot }})^{3}\Upsilon ,  \label{Gamma} \\
\phi _{1}^{*0} &=&(T^{\,{\large \cdot }})^{2}\Phi _{1},  \label{phi1} \\
\phi _{2}^{*0} &=&T^{\,{\large \cdot }}\Phi _{2}.  \label{phi2}
\end{eqnarray}

When these expressions are substituted into Eqs.(\ref{m1}), (\ref{m2}) and 
\ref{gr1}), remembering that 
\begin{equation}
L=-\frac{\text{\dh }T}{T^{{\large \cdot }}},  \label{L&}
\end{equation}
their new form becomes 
\begin{eqnarray}
\text{\dh }\Phi _{1}-\frac{\text{\dh }T}{T^{{\large \cdot }}}\Phi _{1}^{%
{\large \cdot }} &=&0,  \label{m1*} \\
\text{\dh }\Phi _{2}-\frac{\text{\dh }T}{T^{{\large \cdot }}}\Phi _{2}^{%
{\large \cdot }} &=&-(T^{\,{\large \cdot }})^{-1}[\Phi _{1}(T^{\,{\large %
\cdot }})^{2}]^{{\large \cdot }},  \label{m2**} \\
\lbrack \text{\dh }\Upsilon -\frac{\text{\dh }T}{T^{{\large \cdot }}}
\Upsilon ^{{\large \cdot }}] &=&2k\frac{\overline{T}^{\,{\large \cdot }}}{T^{%
{\large \cdot }}}\Phi _{1}\overline{\Phi }_{2}.  \label{gr1*}
\end{eqnarray}

It is these three equations that we will solve within our approximation
scheme.

\begin{remark}
Eq.(\ref{m1*}) can be integrated immediately by 
\begin{equation}
\Phi _{1}=\widehat{\Phi }_{1}(T(u,\zeta ,\widetilde{\zeta })),  \label{PHI^}
\end{equation}

with $\widehat{\Phi }_{1}$ an arbitrary function of $T.$ This follows from
regularity on the sphere and the fact that Eq.(\ref{m1*}) defines a CR
function.
\end{remark}

The main difficulty lies in finding or calculating the $T(u,\zeta ,%
\widetilde{\zeta })$ from the assumed form of the UCF, $X(\tau ,\zeta ,%
\widetilde{\zeta }),$ i.e., from Eq$.(\ref{expansion**}),$

\begin{equation}
u=X(\tau ,\zeta ,\widetilde{\zeta })=\frac{1}{\sqrt{2}}\xi ^{0}(\tau
)Y_{0}^{0}(\zeta ,\widetilde{\zeta })-\frac{1}{2}\xi ^{i}(\tau
)Y_{1i}^{0}(\zeta ,\widetilde{\zeta })+\xi ^{ij}(\tau )Y_{1ij}^{0}(\zeta ,%
\widetilde{\zeta })+....,  \label{expansion***}
\end{equation}
All the relevant functions are assumed to be analytically extended a short
way into $\frak{I}_{C}^{+}.$

$T(u,\zeta ,\widetilde{\zeta })$ can be approximated in the following
manner: From the unit velocity vector $v^{a}(\tau )=\xi ^{a}(\tau )^{\prime
}=(v^{0}(\tau ),v^{i}(\tau ))$ with a slow motion approximation

\begin{equation}
v^{0}(\tau )=\sqrt{1+v^{i\,2}}\cong 1+\frac{1}{2}v^{i\,2},  \label{slow}
\end{equation}
we have 
\begin{equation}
\xi ^{0}(\tau )=\int v^{0}(\tau )d\tau =\tau +\delta \xi ^{0}(\tau )\cong
\tau +\frac{1}{2}\int v^{i\,2}d\tau +\xi _{0}^{0}.  \label{xi_0}
\end{equation}
The Eq.(\ref{expansion***}) can then be written as 
\[
u=\frac{\tau +\delta \xi ^{0}(\tau )}{\sqrt{2}}-\frac{1}{2}\xi ^{i}(\tau
)Y_{1i}^{0}(\zeta ,\widetilde{\zeta })+\xi ^{ij}(\tau )Y_{1ij}^{0}(\zeta ,%
\widetilde{\zeta })+..., 
\]
or as 
\begin{eqnarray}
\sqrt{2}u &=&\tau +\delta \xi ^{0}(\tau )-\frac{\sqrt{2}}{2}\xi ^{i}(\tau
)Y_{1i}^{0}(\zeta ,\widetilde{\zeta })+\sqrt{2}\xi ^{ij}(\tau
)Y_{1ij}^{0}(\zeta ,\widetilde{\zeta })+...,  \label{tau} \\
\tau &=&\sqrt{2}u-\delta \xi ^{0}(\tau )+\frac{\sqrt{2}}{2}\xi ^{i}(\tau
)Y_{1i}^{0}(\zeta ,\widetilde{\zeta })-\sqrt{2}\xi ^{ij}(\tau
)Y_{1ij}^{0}(\zeta ,\widetilde{\zeta })+...,  \nonumber \\
\tau &=&\sqrt{2}u+F(\tau ,\zeta ,\widetilde{\zeta }),  \nonumber
\end{eqnarray}
with

\begin{equation}
F(\tau ,\zeta ,\widetilde{\zeta })=-\delta \xi ^{0}(\tau )+\frac{\sqrt{2}}{2}
\xi ^{i}(\tau )Y_{1i}^{0}(\zeta ,\widetilde{\zeta })-\sqrt{2}\xi ^{ij}(\tau
)Y_{1ij}^{0}(\zeta ,\widetilde{\zeta })+.  \label{F}
\end{equation}

The idea is to iterate Eq.(\ref{tau}).

Taking 
\begin{equation}
\tau =\sqrt{2}u\equiv u_{r},  \label{zeroth}
\end{equation}
as the zeroth iterate (see Eq.(\ref{sqrt2})), we have for the first iterate
that 
\begin{eqnarray}
\tau &=&\sqrt{2}u+F(\sqrt{2}u,\zeta ,\widetilde{\zeta }),  \label{1st} \\
\tau &=&u_{r}+\frac{\sqrt{2}}{2}\xi ^{i}(u_{r})Y_{1i}^{0}(\zeta ,\widetilde{
\zeta })-\sqrt{2}\xi ^{ij}(u_{r})Y_{1ij}^{0}(\zeta ,\widetilde{\zeta }), 
\nonumber
\end{eqnarray}
and for the second iterate 
\begin{equation}
\tau =T(u,\zeta ,\widetilde{\zeta })=u_{r}+F{\large (}u_{r}+F(u_{r},\zeta ,%
\overline{\zeta }),\zeta ,\overline{\zeta }{\large )}\text{ }\approx \text{ }%
u_{r}+F+F\partial _{u_{r}}F.  \label{2ndIt}
\end{equation}

We then find that 
\begin{equation}
T^{\,{\large \cdot }}=\sqrt{2}\partial _{u_{r}}T=\sqrt{2}(1+F^{\prime
}+(F^{\prime })^{2}+FF^{\prime \prime }),  \label{Tdot*}
\end{equation}
or expanded and using the Clebsch-Gordon products, 
\begin{mathletters}
\begin{eqnarray}
T^{\,\,{\large \cdot }} &=&\sqrt{2}{\large (}1+\frac{1}{3}\{v^{i}\xi
^{i}\}^{\prime }-\delta v_{0}+\frac{48}{5}[\xi ^{ij}v^{ij\,\prime
}+v^{ij}v^{ij}]  \label{Tdot^} \\
&&+[\frac{\sqrt{2}}{2}v^{i}-\frac{12}{5}\{v^{ij\,\prime }\xi
^{j}+v^{j\,\prime }\xi ^{ij}+2v^{ij}v^{j}\}]Y_{1i}^{0}  \nonumber \\
&&+\frac{1}{6}\{v^{i\,\prime }\xi ^{k}+v^{i}v^{k}-6\sqrt{2}v^{ki}+\frac{288}{
7}[\xi ^{kl}v^{il\,\prime }+v^{kl}v^{il}]\}Y_{2ki}^{0}{\large ).}  \nonumber
\end{eqnarray}

\begin{remark}
We have used the derivative notion $\partial _{u}=$ ($^{\,\,\cdot }$) and $%
\partial _{\tau }=$ ($^{\prime }$). To avoid a plethora of notations we also
use $\partial _{u_{r}}=$ ($^{\prime }$) since the functions of $u_{r}$
(after expansions) are always \textit{functionally the same} as those of $%
\tau .$ Which variable, $\tau $ or $u_{r},$ is being used is clear from the
context. Furthermore we have $\partial _{u}=$ ($^{\,\,\cdot })=\sqrt{2}%
\partial _{u_{r}}=\sqrt{2}$ ($^{\prime }$). All the harmonic coefficients in
the remainder of this section are functions only of $u_{r}.$
\end{remark}

Functions just of $T,$ as for example the $\widehat{\Phi }_{1}$ of Eq.(\ref
{PHI^}), can be easily approximated as: 
\end{mathletters}
\begin{eqnarray}
\widehat{\Phi }(T) &=&\widehat{\Phi }{\large (}u_{r}+F(u_{r},\zeta ,%
\overline{\zeta })+F\partial _{u_{r}}F{\large )}  \label{G'} \\
&=&\widehat{\Phi }(u_{r})+(F+FF^{\prime })\widehat{\Phi }^{\prime }, 
\nonumber
\end{eqnarray}

or, when $\widehat{\Phi }$ is already first order, 
\begin{equation}
\widehat{\Phi }(T)=\widehat{\Phi }(u_{r})+F\widehat{\Phi }^{\prime }.
\label{G}
\end{equation}

We will take $\widehat{\Phi }_{1}(T),$ with $q=constant$, to have the form 
\begin{equation}
\Phi _{1}=\widehat{\Phi }_{1}(T)=\frac{q}{2}+\delta \Phi _{1}(T).
\label{PHI^*}
\end{equation}

The quantities $\Upsilon ,$ $\Phi _{1}$ and $\Phi _{2}$ are now assumed to
have the harmonic expansion given by 
\begin{eqnarray}
\Phi _{1} &=&\Phi _{1}^{0}+\Phi _{1}^{i}Y_{1i}^{0}+\Phi _{1}^{ij}Y_{2ij}^{0},
\label{PHI_1exp} \\
\Phi _{2} &=&\Phi _{2}^{i}Y_{1i}^{-1}+\Phi _{2}^{ij}Y_{2ij}^{-1},
\label{PHI_2exp} \\
\Upsilon &=&\Upsilon ^{0}+\Upsilon ^{i}Y_{1i}^{0}+\Upsilon ^{ij}Y_{2ij}^{0}.
\label{Gammaexp}
\end{eqnarray}

By substituting these into the field equations, Eqs.(\ref{m1*}),\ (\ref{m2**}%
) and (\ref{gr1*}), (or Eq.(\ref{G}) for $\Phi _{1}$) and using Eqs.(\ref
{2ndIt}) and (\ref{Tdot^}), after very lengthy calculations with frequent
use of the Clebsch-Gordon expansions we obtain the solution to the new
version of the Maxwell equations 
\begin{eqnarray}
\phi _{1}^{*0} &=&(T^{\,{\large \cdot }})^{2}\Phi _{1},  \label{PHI_1} \\
\Phi _{1}(T) &=&\frac{q}{2}+\delta \Phi _{1}(T)=\frac{q}{2}(1-\frac{144}{5}
v^{ij2}),
\end{eqnarray}
and 
\begin{eqnarray}
\phi _{2}^{*0} &=&(T^{\,{\large \cdot }})\Phi _{2},  \label{PHI_2} \\
\Phi _{2} &=&\Phi ^{i}Y_{1i}^{-1}+\Phi ^{ij}Y_{2ij}^{-1}, \\
\Phi ^{i} &=&-\sqrt{2}qv^{i\prime }+\frac{12}{5}q[v^{ij\prime \prime }\xi
^{j}+6v^{ij\prime }v^{j}-v^{j\prime \prime }\xi ^{ij}+6v^{j\prime }v^{ij}]
\label{PHI^e} \\
&&-i\sqrt{2}\frac{q}{2}{\large \{}\xi ^{l}v^{k\prime \prime }-\frac{24(4)}{5}
\xi ^{kj}v^{lj\prime \prime }{\large \}}\epsilon _{lki},  \nonumber \\
\Phi ^{il} &=&2\sqrt{2}qv^{li\prime }+2i\sqrt{2}q{\small SymTF}(\xi
^{ij}v^{k\prime \prime }-\frac{1}{3}v^{ij\prime \prime }\xi ^{k})\epsilon
_{jkl}  \label{PHI^ij} \\
&&-\frac{q}{6}{\small SymTF}{\large \{}3v^{i\prime \prime }\xi
^{l}+6v^{i\prime }v^{l}+\frac{288}{7}(6v^{ls}v^{is\prime }+\xi
^{ls}v^{is\prime \prime }){\large \}.}  \nonumber
\end{eqnarray}

$SymTF$ means take the symmetric trace-free part.

When these results are substituted into the gravitational equation,

\begin{equation}
\lbrack \text{\dh }\Upsilon -\frac{\text{\dh }T}{T^{{\large \cdot }}}%
\Upsilon ^{{\large \cdot }}]=2k\frac{\overline{T}^{\,{\large \cdot }}}{T^{%
{\large \cdot }}}\Phi _{1}\overline{\Phi }_{2},  \label{GAMMA}
\end{equation}
we see that the quadratic part of $\Phi _{1}$ drops out and plays no further
role so that we could have written 
\begin{equation}
\Phi _{1}(T)=\frac{q}{2}.  \label{PHI_1*}
\end{equation}

Finally we have the solution of Eq.(\ref{GAMMA}) 
\begin{eqnarray}
\psi _{2}^{*0} &=&-(T^{\,{\large \cdot }})^{3}\Upsilon ,  \label{GAMMA*} \\
\Upsilon &=&-\Upsilon _{0}+\Upsilon ^{i}Y_{1i}^{0}+\Upsilon ^{ij}Y_{2ij}^{0},
\label{GAMMAexp}
\end{eqnarray}
where 
\begin{eqnarray}
\Upsilon ^{e} &=&-k\frac{q^{2}}{4}{\large \{}-2\sqrt{2}\overline{v}^{e\prime
}+\frac{8}{5}{\large [}(3\overline{\xi }^{j}-\xi ^{j})\overline{v}^{ej\prime
\prime }+3\overline{v}^{ej\prime }(7\overline{v}^{j}-v^{j})  \label{GAMMAe}
\\
&&-3(\overline{\xi }^{ej}-3{\xi }^{ej})\overline{v}^{j\prime \prime }+3%
\overline{v}^{j\prime }(5\overline{v}^{ej}+v^{ej}){\large ]\}}  \nonumber \\
&&-k\frac{q^{2}}{4}i\sqrt{2}{\large \{}(\overline{\xi }^{i}-\xi ^{i})%
\overline{v}^{k\prime }-\frac{24(4)}{5}(\overline{\xi }^{kj}-\xi ^{kj})%
\overline{v}^{ij\prime }{\large \}}^{\prime }\epsilon _{ike},  \nonumber
\end{eqnarray}

\begin{eqnarray}
\Upsilon ^{il} &=&-\frac{1}{3}k\frac{q^{2}}{4}{\large \{}4\sqrt{2}\overline{v%
}^{li\prime }-\frac{2}{6}{\small SymTF}{\large [}3(\overline{\xi }^{l}+\xi
^{l})\overline{v}^{i\prime \prime }+3\overline{v}^{i\prime }(3\overline{v}
^{l}-v^{l})  \label{GAMMAil} \\
&&+\frac{288}{7}[\overline{v}^{is\prime }(7\overline{v}^{ls}-v^{ls})+(%
\overline{\xi }^{ls}+\xi ^{ls})\overline{v}^{is\prime \prime }]{\large ]\}} 
\nonumber \\
&&+\frac{i\sqrt{2}kq^{2}}{3}{\small SymTF}{\large [}(\overline{\xi }
^{ij}-\xi ^{ij})\overline{v}^{k\prime }-\frac{1}{3}(\overline{\xi }^{k}-\xi
^{k})\overline{v}^{ij\prime }{\large ]}^{\prime }\epsilon _{jkl}.  \nonumber
\end{eqnarray}

Before we go on to the evolution equation, Eq.(\ref{gr2}), we remark that
these calculations were extremely long and rather ugly. They however were
checked by doing them other ways; via the $L$ version, Eqs.(\ref{m1}), (\ref
{m2}) and (\ref{gr1}) and then partially by the RTMaxwell version\cite{RTM}.

\subsection{The Bondi Mass Aspect}

In this section we calculate the Bondi mass aspect, Eq.(\ref{gr3}), 
\begin{equation}
\Psi =\overline{\Psi }=\psi _{2}^{*0}+2L\text{\dh }(\overline{\sigma })^{%
{\large \cdot }\,\,}+L^{2}(\overline{\sigma })^{{\large \cdot \cdot }}+\text{
\dh }^{2}\overline{\sigma }+\sigma (\overline{\sigma })^{{\large \cdot }},
\label{MA}
\end{equation}
and then, from the $l=(0,1)$ harmonics, extract the energy-momentum four
vector. Again the calculations are long and complicated with frequent uses
of the Clebsch-Gordon expansions.

First, defining the harmonic coefficients by, 
\begin{equation}
\psi _{2}^{*0}=\Gamma _{0}+\Gamma ^{i}Y_{1i}^{0}+\Gamma ^{ij}Y_{2ij}^{0},
\label{psi2*}
\end{equation}
and using 
\begin{equation}
\psi _{2}^{*0}=[-\Upsilon _{0}+\Upsilon ^{i}Y_{1i}^{0}+\Upsilon
^{ij}Y_{2ij}^{0}]T^{\,\,{\large \cdot 3}},  \label{psi2**}
\end{equation}
we obtain, from Eqs.(\ref{Tdot^}), (\ref{GAMMA*}) and (\ref{GAMMAexp}), that 
\begin{eqnarray}
\Gamma _{0} &=&-2\sqrt{2}\Upsilon _{0}[1+\frac{1}{2}v^{i}v^{i}+v^{i\prime
}\xi ^{i}+\frac{144}{5}(\xi ^{ij}v^{ij\prime }+2v^{ij}v^{ij})]  \label{Gamm0}
\\
&&+2\sqrt{2}kq^{2}[\overline{v}^{i\prime }v^{i}+\frac{48}{5}\overline{v}%
^{ij\prime }v^{ij}],  \nonumber
\end{eqnarray}

\begin{eqnarray}
\Gamma ^{l} &=&-6\Upsilon _{0}v^{l}+2kq^{2}\overline{v}^{l\prime }+\Upsilon
_{0}\frac{72\sqrt{2}}{5}(v^{lj\prime }\xi ^{j}+v^{j\prime }\xi
^{lj}+4v^{lj}v^{j}) \\
&&-kq^{2}\frac{4\sqrt{2}}{5}[(3\overline{\xi }^{j}-\xi ^{j})\overline{v}
^{lj\prime \prime }+3\overline{v}^{lj\prime }(7\overline{v}^{j}+v^{j})-3(%
\overline{\xi }^{lj}-3{\xi }^{lj})\overline{v}^{j\prime \prime }+3\overline{v%
}^{j\prime }(5\overline{v}^{lj}+7v^{lj})]  \nonumber \\
&&-kq^{2}i[(\overline{\xi }^{i}-\xi ^{i})\overline{v}^{k\prime }-\frac{96}{5}
(\overline{\xi }^{kj}-\xi ^{kj})\overline{v}^{ij\prime }]^{\prime }\epsilon
_{ikl},  \nonumber
\end{eqnarray}

\begin{eqnarray}
\Gamma ^{ik} &=&12\Upsilon _{0}v^{ki}-\frac{4kq^{2}}{3}\overline{v}
^{ki\prime }  \label{Gammaij} \\
&&-\Upsilon _{0}\sqrt{2}{\small SymTF}[v^{i\prime }\xi ^{k}+2v^{i}v^{k}+%
\frac{288}{7}(\xi ^{kl}v^{il\prime }+2v^{kl}v^{il})]  \nonumber \\
&&+\frac{\sqrt{2}kq^{2}}{6}{\small SymTF}{\large \{}(\overline{\xi }^{k}+\xi
^{k})\overline{v}^{i\prime \prime }+\overline{v}^{i\prime }(3\overline{v}
^{k}+5v^{k})+\frac{96}{7}[\overline{v}^{is\prime }(7\overline{v}%
^{ks}+5v^{ks})  \nonumber \\
&&+(\overline{\xi }^{ks}+\xi ^{ks})\overline{v}^{is\prime \prime }]+\frac{4i%
\sqrt{2}}{3}[(\overline{\xi }^{ij}-\xi ^{ij})\overline{v}^{s\prime }-\frac{1%
}{3}(\overline{\xi }^{s}-\xi ^{s})\overline{v}^{ij\prime }]^{\prime
}\epsilon _{jsk}{\large \}}.  \nonumber
\end{eqnarray}

The remaining terms in the mass aspect, Eq.(\ref{MA}), i.e., 
\[
2L\text{\dh }(\overline{\sigma })^{\,\,{\large \cdot }}+L^{2}(\overline{
\sigma })^{{\large \cdot \cdot }}+\text{\dh }^{2}\overline{\sigma }+\sigma (%
\overline{\sigma })^{{\large \cdot }}, 
\]
are calculated from Eqs.(\ref{SIGMA}), (\ref{L&}), (\ref{2ndIt}) and (\ref
{Tdot^}).

Using this result and the expression for $\psi _{2}^{*0}$, we obtain 
\begin{equation}
\Psi =\Psi ^{(0)}+\Psi ^{(i)}Y_{1i}^{0}+\Psi ^{(ij)}Y_{2ij}^{0},  \label{MA*}
\end{equation}
with 
\begin{eqnarray}
\Psi ^{(0)} &=&-2\sqrt{2}\{\Upsilon _{0}[1+\frac{1}{2}v^{i}v^{i}+v^{i\prime
}\xi ^{i}+\frac{144}{5}(\xi ^{ij}v^{ij\prime }+2v^{ij}v^{ij})]+\frac{288}{5}%
\xi ^{ij}\overline{v}^{ij}  \label{MA0} \\
&&-kq^{2}[\overline{v}^{i\prime }v^{i}+\frac{48}{5}\overline{v}^{ij\prime
}v^{ij}]\},  \nonumber
\end{eqnarray}

\begin{eqnarray}
\Psi ^{(l)} &=&-6{\large \{}\Upsilon _{0}v^{l}-\frac{1}{3}kq^{2}\overline{v}
^{l\prime }-\Upsilon _{0}\frac{12\sqrt{2}}{5}(v^{lj\prime }\xi
^{j}+v^{j\prime }\xi ^{lj}+4v^{lj}v^{j})-\frac{24\sqrt{2}}{5}\overline{v}
^{lj}\xi ^{j}  \label{MAe} \\
&&+kq^{2}\frac{2\sqrt{2}}{15}[(3\overline{\xi }^{j}-\xi ^{j})\overline{v}%
^{lj\prime \prime }+3\overline{v}^{lj\prime }(7\overline{v}^{j}+v^{j})-3(%
\overline{\xi }^{lj}-3{\xi }^{lj})\overline{v}^{j\prime \prime }  \nonumber
\\
&&+3\overline{v}^{j\prime }(5\overline{v}^{lj}+7v^{lj})]+\frac{1}{6}ikq^{2}[(%
\overline{\xi }^{i}-\xi ^{i})\overline{v}^{k\prime }-\frac{96}{5}(\overline{
\xi }^{kj}-\xi ^{kj})\overline{v}^{ij\prime }]^{\prime }\epsilon _{ikl}%
{\large \},}  \nonumber
\end{eqnarray}

\begin{eqnarray}
\Psi ^{(ik)} &=&12\Upsilon _{0}v^{ki}-\frac{4kq^{2}}{3}\overline{v}
^{ki\prime }+24\overline{\xi }^{ik}  \label{MAij} \\
&&-\Upsilon _{0}\sqrt{2}{\small SymTF}[v^{i\prime }\xi ^{k}+2v^{i}v^{k}+%
\frac{288}{7}(\xi ^{kl}v^{il\prime }+2v^{kl}v^{il})]  \nonumber \\
&&+\frac{\sqrt{2}kq^{2}}{6}{\small SymTF}{\large \{}(\overline{\xi }^{k}+\xi
^{k})\overline{v}^{i\prime \prime }+\overline{v}^{i\prime }(3\overline{v}
^{k}+5v^{k})  \nonumber \\
&&+\frac{96}{7}[\overline{v}^{is\prime }(7\overline{v}^{ks}+5v^{ks})+(%
\overline{\xi }^{ks}+\xi ^{ks})\overline{v}^{is\prime \prime }]  \nonumber \\
&&+\frac{4i\sqrt{2}}{3}[(\overline{\xi }^{ij}-\xi ^{ij})\overline{v}
^{s\prime }-\frac{1}{3}(\overline{\xi }^{s}-\xi ^{s})\overline{v}^{ij\prime }%
]^{\prime }\epsilon _{jsk}{\large \}}  \nonumber \\
&&-{\small SymTF}[\frac{(24)^{2}\sqrt{2}}{7}\overline{v}^{is}(2\xi ^{ks}-%
\overline{\xi }^{ks})-16i\overline{v}^{in}(\overline{\xi }^{s}-\xi
^{s})\epsilon _{snk}],  \nonumber
\end{eqnarray}

Finally, the condition 
\begin{equation}
\Psi =\overline{\Psi },  \label{MA=MAbar}
\end{equation}
yields the \textit{reality conditions} to be imposed on the harmonic
components 
\begin{eqnarray}
\Psi ^{(0)}-\overline{\Psi }^{(0)} &=&0,  \label{reality1} \\
\Psi ^{(e)}-\overline{\Psi }^{(e)} &=&0,  \label{reality2} \\
\Psi ^{(ik)}-\overline{\Psi }^{(ik)} &=&0.  \label{reality3}
\end{eqnarray}

\section{The Bondi Energy-Momentum Four-Vector}

The relationship of the mass aspect, $\Psi _{real}\equiv \Psi =\frac{\Psi +%
\overline{\Psi }}{2},$ to the Bondi energy-momentum vector, 
\begin{equation}
P^{a}=(Mc,P^{i}),  \label{P^a}
\end{equation}
is\cite{Sachs,RP,NT,KNO,KNO2,KN},

\begin{equation}
\Psi _{real}=-\frac{2\sqrt{2}G}{c^{2}}M-\frac{6G}{c^{3}}P^{i}Y_{1i}^{0}+\Psi
^{il}Y_{2il}^{0}.  \label{mass.asp.=>m,p}
\end{equation}
By comparing with Eqs.(\ref{mass.asp.=>m,p}) with the final form of the mass
aspect, Eq.(\ref{MA*}), - (which is now \textit{real) - }the mass, $M$ and
momentum, $P^{i}$ can be found.

Using the relation 
\[
\xi ^{i}=\xi _{R}^{i}+i\xi _{I}^{i},\text{ etc.,} 
\]
leads to the kinematic relations between mass and momentum with the
position, velocity and acceleration, quantities defined from the UCF, Eq.(%
\ref{expansion***}), i.e.,

\begin{eqnarray}
M &=&M_{0}\{1+\frac{1}{2c^{2}}(v_{R}^{i}v_{R}^{i}-v_{I}^{i}v_{I}^{i})+\frac{1%
}{c^{2}}(v_{R}^{i\prime }\xi _{R}^{i}-v_{I}^{i\prime }\xi _{I}^{i})
\label{M*} \\
&&+\frac{144}{5c^{2}}[v_{R}^{ij\prime }\xi _{R}^{ij}-v_{I}^{ij\prime }\xi
_{I}^{ij}+2(v_{R}^{ij}v_{R}^{ij}-v_{I}^{ij}v_{I}^{ij})]\}  \nonumber \\
&&+\frac{288c}{5G}(\xi _{R}^{ij}v_{R}^{ij}+\xi _{I}^{ij}v_{I}^{ij})-\frac{%
2q^{2}}{c^{5}}\{v_{R}^{i\prime }v_{R}^{i}+v_{I}^{i\prime }v_{I}^{i}+\frac{48%
}{5}(v_{R}^{ij\prime }v_{R}^{ij}+v_{I}^{ij\prime }v_{I}^{ij})\},  \nonumber
\end{eqnarray}

\begin{eqnarray}
P^{e} &=&M_{0}v_{R}^{e}-\frac{2q^{2}}{3c^{3}}v_{R}^{e\prime }-\frac{12\sqrt{2%
}M_{0}}{5c}[v_{R}^{ej\prime }\xi _{R}^{j}-v_{I}^{ej\prime }\xi
_{I}^{j}+v_{R}^{j\prime }\xi _{R}^{ej}-v_{I}^{j\prime }\xi
_{I}^{ej}+4(v_{R}^{ej}v_{R}^{j}-v_{I}^{ej}v_{I}^{j})]  \label{P^i} \\
&&-\frac{24\sqrt{2}c^{2}}{5G}[v_{R}^{ej}\xi _{R}^{j}+v_{I}^{ej}\xi _{I}^{j}]+%
\frac{8\sqrt{2}q^{2}}{15c^{4}}[\xi _{R}^{j}v_{R}^{ej\prime \prime }-2\xi
_{I}^{j}v_{I}^{ej\prime \prime }+12v_{R}^{ej\prime
}v_{R}^{j}-9v_{I}^{ej\prime }v_{I}^{j}  \nonumber \\
&&+3\xi _{R}^{ej}v_{R}^{j\prime \prime }+6\xi _{I}^{ej}v_{I}^{j\prime \prime
}+18v_{R}^{j\prime }v_{R}^{ej}+3v_{I}^{j\prime }v_{I}^{ej}]+\frac{2q^{2}}{%
3c^{4}}[\xi _{I}^{i}v_{R}^{k\prime }-\frac{96}{5}\xi
_{I}^{kj}v_{R}^{ij\prime }]^{\prime }\epsilon _{ike},  \nonumber
\end{eqnarray}
and

\begin{eqnarray}
\Psi ^{(ik)} &=&\frac{12GM_{0}}{c^{3}}v_{R}^{ki}-\frac{8Gq^{2}}{3c^{6}}
v_{R}^{ki\prime }+24\xi _{R}^{ik}  \label{PSI^ij} \\
&&-\frac{\sqrt{2}GM_{0}}{c^{4}}{\small SymTF}[v_{R}^{i\prime }\xi
_{R}^{k}-v_{I}^{i\prime }\xi
_{I}^{k}+2v_{R}^{i}v_{R}^{k}-2v_{I}^{i}v_{I}^{k}+\frac{288}{7}(\xi
_{R}^{kl}v_{R}^{il\prime }-\xi _{I}^{kl}v_{I}^{il\prime }  \nonumber \\
&&+2v_{R}^{kl}v_{R}^{il}-2v_{I}^{kl}v_{I}^{il})]  \nonumber \\
&&+\frac{2\sqrt{2}Gq^{2}}{3c^{7}}{\small SymTF}[\xi _{R}^{k}v_{R}^{i\prime
\prime }+4v_{R}^{i\prime }v_{R}^{k}+v_{I}^{i\prime }v_{I}^{k}+\frac{96}{7}%
(6v_{R}^{is\prime }v_{R}^{ks}-v_{I}^{is\prime }v_{I}^{ks}+\xi
_{R}^{ks}v_{R}^{is\prime \prime })  \nonumber \\
&&+\frac{4\sqrt{2}}{3}(\xi _{I}^{ij}v_{R}^{s\prime }-\frac{1}{3}\xi
_{I}^{s}v_{R}^{ij\prime })^{\prime }\epsilon _{jsk}]  \nonumber \\
&&-\frac{(24)^{2}\sqrt{2}}{7c}{\small SymTF}[v_{R}^{is}\xi
_{R}^{ks}+3v_{I}^{is}\xi _{I}^{ks}-32\xi _{I}^{s}v_{R}^{in}\epsilon _{snk}].
\nonumber
\end{eqnarray}

The reality conditions, Eqs.(\ref{reality1}), (\ref{reality2}) and (\ref
{reality3}) become 
\begin{eqnarray}
0 &=&M_{0}[\frac{1}{c^{2}}v_{R}^{i}v_{I}^{i}+\frac{1}{c^{2}}(v_{R}^{i\prime
}\xi _{I}^{i}+v_{I}^{i\prime }\xi _{R}^{i})+\frac{144}{5c^{2}}
(v_{R}^{ij\prime }\xi _{I}^{ij}+v_{I}^{ij\prime }\xi
_{R}^{ij}+4v_{R}^{ij}v_{I}^{ij})]  \label{reality1*} \\
&&+\frac{288c}{5G}(\xi _{I}^{ij}v_{R}^{ij}-\xi _{R}^{ij}v_{I}^{ij})-\frac{%
2q^{2}}{c^{5}}[v_{R}^{i\prime }v_{I}^{i}-v_{I}^{i\prime }v_{R}^{i}+\frac{48}{%
5}(v_{R}^{ij\prime }v_{I}^{ij}-v_{I}^{ij\prime }v_{R}^{ij})],  \nonumber
\end{eqnarray}
\begin{eqnarray}
0 &=&M_{0}v_{I}^{e}+\frac{2q^{2}}{3c^{3}}v_{I}^{e\prime }-\frac{12\sqrt{2}
M_{0}}{5c}[v_{R}^{ej\prime }\xi _{I}^{j}+v_{I}^{ej\prime }\xi
_{R}^{j}+v_{R}^{j\prime }\xi _{I}^{ej}+v_{I}^{j\prime }\xi
_{R}^{ej}+4(v_{R}^{ej}v_{I}^{j}+v_{I}^{ej}v_{R}^{j})]  \label{reality2*} \\
&&-\frac{24\sqrt{2}c^{2}}{5G}[v_{R}^{ej}\xi _{I}^{j}-v_{I}^{ej}\xi _{R}^{j}]-%
\frac{8\sqrt{2}q^{2}}{15c^{4}}[\xi _{R}^{j}v_{I}^{ej\prime \prime }+2\xi
_{I}^{j}v_{R}^{ej\prime \prime }+9v_{R}^{ej\prime
}v_{I}^{j}+12v_{I}^{ej\prime }v_{R}^{j}  \nonumber \\
&&-6\xi _{I}^{ej}v_{R}^{j\prime \prime }+3\xi _{R}^{ej}v_{I}^{j\prime \prime
}-3v_{R}^{j\prime }v_{I}^{ej}+18v_{I}^{j\prime }v_{R}^{ej}]-\frac{2q^{2}}{
3c^{4}}[\xi _{I}^{i}v_{I}^{k\prime }-\frac{96}{5}\xi
_{I}^{kj}v_{I}^{ij\prime }]^{\prime }\epsilon _{ike}.  \nonumber
\end{eqnarray}

and

\begin{eqnarray}
0 &=&\frac{12GM_{0}}{c^{3}}v_{I}^{ki}+\frac{8Gq^{2}}{3c^{6}}v_{I}^{ki\prime
}-24\xi _{I}^{ik}  \nonumber \\
&&-\frac{\sqrt{2}GM_{0}}{c^{4}}{\small SymTF}[v_{R}^{i\prime }\xi
_{I}^{k}+v_{I}^{i\prime }\xi
_{R}^{k}+2v_{R}^{i}v_{I}^{k}+2v_{I}^{i}v_{R}^{k}+\frac{288}{7}(\xi
_{R}^{kl}v_{I}^{il\prime }+\xi _{I}^{kl}v_{R}^{il\prime
}+2v_{R}^{kl}v_{I}^{il}+2v_{I}^{kl}v_{R}^{il}]  \nonumber \\
&&-\frac{2\sqrt{2}Gq^{2}}{3c^{7}}{\small SymTF}[\xi _{R}^{k}v_{I}^{i\prime
\prime }-v_{R}^{i\prime }v_{I}^{k}+4v_{I}^{i\prime }v_{R}^{k}+\frac{96}{7}%
(v_{R}^{is\prime }v_{I}^{ks}+6v_{I}^{is\prime }v_{R}^{ks}+\xi
_{R}^{ks}v_{I}^{is\prime \prime })  \nonumber \\
&&+\frac{4\sqrt{2}}{3}(\xi _{I}^{ij}v_{I}^{s\prime }-\frac{1}{3}\xi
_{I}^{s}v_{I}^{ij\prime })^{\prime }\epsilon _{jsk}]  \label{reality3*} \\
&&-\frac{(24)^{2}\sqrt{2}}{7c}{\small SymTF}[3v_{R}^{is}\xi
_{I}^{ks}-v_{I}^{is}\xi _{R}^{ks}+32\xi _{I}^{s}v_{I}^{in}\epsilon _{snk}]. 
\nonumber
\end{eqnarray}

These `kinematic' equations are of sufficient complexity that understanding
of them is difficult or even problematic. It appears that perhaps the best
way to proceed is to leave out for the time being some of the details -
mainly the affect of the $l=2$ terms, e.g., .($\xi _{R}^{ij},v_{R}^{ij},\xi
_{I}^{kj},v_{I}^{ij}$) and their derivatives.

We then obtain for the mass 
\begin{equation}
M=M_{0}\{1+\frac{1}{2c^{2}}(v_{R}^{i}v_{R}^{i}-v_{I}^{i}v_{I}^{i})+\frac{1}{
c^{2}}(v_{R}^{i\prime }\xi _{R}^{i}-v_{I}^{i\prime }\xi _{I}^{i})\}-\frac{%
2q^{2}}{c^{5}}\{v_{R}^{i\prime }v_{R}^{i}+v_{I}^{i\prime }v_{I}^{i}\},
\label{M***}
\end{equation}
the rest mass and kinetic energy term plus a series of unusual quadratic
terms that do not appear to have any standard counter parts. If - as we
argue later - we identify 
\[
d_{E}^{i}=q\xi _{R}^{i}\text{ and }d_{M}^{i}=q\xi _{I}^{i}, 
\]
as the electric and magnetic dipole moments and then arguing from the Kerr
and charged Kerr solutions that we identify 
\begin{eqnarray*}
S^{i} &=&M_{0}c\xi _{I}^{i}, \\
S^{i\prime } &=&M_{0}cv_{I}^{i},
\end{eqnarray*}
as the spin dipole moment and its time derivative, we can interpret the
extra terms in the mass as the interactions of these moments with themselves
and each other and write: 
\begin{eqnarray}
M &=&M_{0}\{1+\frac{1}{2c^{2}}(v_{R}^{i}v_{R}^{i}-\frac{1}{M_{0}c^{2}}
S^{i\prime }S^{i\prime })+\frac{1}{c^{2}}(v_{R}^{i\prime }\xi _{R}^{i}-\frac{
1}{M_{0}c^{2}}S^{i\prime \prime }S^{i})\}  \label{M****} \\
&&-\frac{2}{c^{5}}\{d_{E}^{i\prime \prime }d_{E}^{i\prime }+d_{M}^{i\prime
\prime }d_{M}^{i\prime }\}.  \nonumber
\end{eqnarray}

.The linear momentum expression becomes 
\begin{equation}
P^{e}=M_{0}v_{R}^{e}-\frac{2q^{2}}{3c^{3}}v_{R}^{e\prime }+\frac{2}{3c^{4}}[
d_{M}^{i}d_{E}^{j\prime \prime }]^{\prime }\epsilon _{ike},  \nonumber
\end{equation}
with the first term being the standard $Mv$, the second the contribution
from radiation reaction and the last term is new.

The first and second reality conditions become \thinspace

\begin{equation}
0=M_{0}[v_{R}^{i}v_{I}^{i}+(v_{R}^{i\prime }\xi _{I}^{i}+v_{I}^{i\prime }\xi
_{R}^{i})]-\frac{2q^{2}}{c^{3}}[v_{R}^{i\prime }v_{I}^{i}-v_{I}^{i\prime
}v_{R}^{i}],  \label{first}
\end{equation}

\begin{equation}
0=M_{0}v_{I}^{e}+\frac{2q^{2}}{3c^{3}}v_{I}^{e\prime }-\frac{2q^{2}}{3c^{4}}[
\xi _{I}^{i}v_{I}^{k\prime }]^{\prime }\epsilon _{ike}.  \label{second}
\end{equation}

We see from Eq.(\ref{second}) that when the charge vanishes $v_{I}^{e}=0$
and the spin S$^{i}$ is constant. From Eq.(\ref{first}), $S^{i}$ is normal
to $v_{R}^{i\prime }.$ When $q\neq 0,$ $v_{I}^{e}$ exponentially decreases
and S$^{i}$ is asymptotically normal to $v_{R}^{i\prime }.$

The third reality condition has the $\xi _{I}^{ij}$ diverge exponentially at
linear order though it is not clear what are the affects of the
non-linearity.

It appears that the reality conditions play the physical role of describing
the evolution of the spin (or so-called `magnetic moments').

We thus have found from the mass aspect and the reality conditions the
following information: (a) there is a kinematic expression for the total
mass, i.e., the bare mass dressed with the kinetic energy and spin-velocity
coupled terms, (b) for the momentum there is the usual kinetic term with the
classical radiation reaction term plus an electric/magnetic dipole
coupling., (c) the first two reality conditions, though not fully
understood, determine the behavior of the spin vector, (d) while the $l=2$
reality condition, also not well understood, determines the imaginary or
spin part of the quadrupole term.

\section{Dynamics: Equations of Motion}

The equations that we have been dealing with have essentially been
relational, i.e., they just giving the relations between the different
geometrical/physical quantities. Essentially all the variables have been
determined in terms of the harmonic coefficients of the UCF and their
derivatives, i.e., ($\xi ^{0},\xi ^{i},\xi ^{ij}$), ($v^{0},v^{i},v^{ij}$),
etc. and the $\Upsilon _{0}.$

The final dynamical equation that determine the evolution of these
quantities is the Bondi equation, (\ref{gr2}), 
\begin{equation}
\Psi ^{{\large \cdot }}=\sigma ^{{\large \cdot }}\overline{\sigma }^{{\large %
\cdot }}+k\phi _{2}^{*0}\overline{\phi }_{2}^{*0}.  \label{PSIdot}
\end{equation}

By the harmonic expansion of $\Psi ,$ Eq.(\ref{mass.asp.=>m,p}), and using
the expressions for $\phi _{2}^{*0}$ and $\sigma ^{\cdot }$ in Eq.(\ref
{PSIdot}), we obtain the mass and momentum loss equations,

\begin{eqnarray}
M^{\prime } &=&-\frac{288c}{5G}(v_{R}^{ij}v_{R}^{ij}+v_{I}^{ij}v_{I}^{ij})-%
\frac{2q^{2}}{3c^{5}}(v_{R}^{i\prime }v_{R}^{i\prime }+v_{I}^{i\prime
}v_{I}^{i\prime })-\frac{32q^{2}}{5c^{5}}(v_{R}^{ij\prime }v_{R}^{ij\prime
}+v_{I}^{ij\prime }v_{I}^{ij\prime }),  \label{m'} \\
P^{e\prime } &=&\frac{192c^{2}}{5G}
(v_{I}^{kj}v_{R}^{ij}-v_{R}^{kj}v_{I}^{ij})\epsilon _{ike}-\frac{q^{2}}{
3c^{4}}(v_{I}^{j\prime }v_{R}^{i\prime }-v_{R}^{j\prime }v_{I}^{i\prime
})\epsilon _{ije}.  \label{P^i'} \\
&&+\frac{8\sqrt{2}q^{2}}{5c^{4}}(v_{R}^{j\prime }v_{R}^{je\prime
}+v_{I}^{j\prime }v_{I}^{je\prime })+\frac{32q^{2}}{15c^{4}}(v_{R}^{kj\prime
}v_{I}^{ij\prime }-v_{I}^{kj\prime }v_{R}^{ij\prime })\epsilon _{ike}. 
\nonumber
\end{eqnarray}

The mass/energy loss equation (\ref{m'}) has the following simple meaning:

\begin{itemize}
\item  If, in the first term on the right side, we make the physical
identification of ($v_{R}^{ij},v_{I}^{ij})$ with the derivatives of
gravitational mass and spin quadrupole moments via 
\begin{equation}
(v_{R}^{ij},v_{I}^{ij})=\frac{G}{12\sqrt{2}c^{4}}(Q_{Mass}^{ij\prime \prime
\prime },Q_{Spin}^{ij\prime \prime \prime }),  \label{quad.moments}
\end{equation}
it is exactly the Bondi gravitational quadrupole energy loss.

\item  For the second term, if we identify, as we mentioned earlier, $q(\xi
_{R}^{i},\xi _{I}^{i})$ with the electric and magnetic dipole moments, it is
precisely both electric and magnetic dipole energy loss.

\item  If, in the third term, we make the identification of $q$($%
v_{R}^{ij},v_{I}^{ij})$ with the derivatives of the electric and magnetic
quadrupole moments via 
\[
q(v_{R}^{ij},v_{I}^{ij})=\frac{1}{24\sqrt{2}c}(D_{Elec}^{ij\prime \prime
},D_{Mag}^{ij\prime \prime }), 
\]
we have the electromagnetic quadrupole energy loss.
\end{itemize}

The momentum loss equation , (\ref{P^i'}), also has a simple physical
interpretation, namely it is the source recoil from the radiated
gravitational and electromagnetic momentum.

Our final task is to substitute our dynamic expressions for the mass and
momentum, Eqs.(\ref{M*}) and (\ref{P^i}, into the mass and momentum loss
equations, (\ref{m'}) and (\ref{P^i'}), obtaining the equations of motion
for what we define as the center of mass,

\begin{eqnarray}
M_{0}v_{R}^{e\prime } &=&\frac{2q^{2}}{3c^{3}}v_{R}^{e\prime \prime }+\frac{%
2q^{2}}{3c^{4}}[v_{I}^{i\prime }v_{R}^{j\prime }-(\xi _{I}^{i}v_{R}^{j\prime
})^{\prime \prime }]\epsilon _{ije}+J^{e},  \label{EqsOfMotionII} \\
J^{e} &=&{\large \{}\frac{12\sqrt{2}M_{0}}{5c}[v_{R}^{ej\prime }\xi
_{R}^{j}-v_{I}^{ej\prime }\xi _{I}^{j}+v_{R}^{j\prime }\xi
_{R}^{ej}-v_{I}^{j\prime }\xi
_{I}^{ej}+4(v_{R}^{ej}v_{R}^{j}-v_{I}^{ej}v_{I}^{j})]  \nonumber \\
&&+\frac{24\sqrt{2}c^{2}}{5G}[v_{R}^{ej}\xi _{R}^{j}+v_{I}^{ej}\xi _{I}^{j}]-%
\frac{8\sqrt{2}q^{2}}{15c^{4}}[\xi _{R}^{j}v_{R}^{ej\prime \prime }-2\xi
_{I}^{j}v_{I}^{ej\prime \prime }+12v_{R}^{ej\prime
}v_{R}^{j}-9v_{I}^{ej\prime }v_{I}^{j}  \nonumber \\
&&+3\xi _{R}^{ej}v_{R}^{j\prime \prime }+6\xi _{I}^{ej}v_{I}^{j\prime \prime
}+18v_{R}^{j\prime }v_{R}^{ej}+3v_{I}^{j\prime }v_{I}^{ej}]+\frac{2q^{2}}{%
3c^{4}}[\frac{96}{5}\xi _{I}^{kj}v_{R}^{ij\prime }]^{\prime }\epsilon _{ike}%
{\large \}}^{\prime }  \nonumber \\
&&+\frac{384c^{2}}{5G}v_{I}^{kj}v_{R}^{ij}\epsilon _{ike}+\frac{8\sqrt{2}
q^{2}}{5c^{4}}(v_{R}^{j\prime }v_{R}^{je\prime }+v_{I}^{j\prime
}v_{I}^{je\prime })+\frac{64q^{2}}{15c^{4}}v_{R}^{kj\prime }v_{I}^{ij\prime
}\epsilon _{ike}  \nonumber
\end{eqnarray}
with the term $J^{e}$ depending on the $l=2$ harmonics. We see a standard
left side, mass times acceleration, equal to the classical radiation
reaction term plus a coupling, (cross-products) between derivatives of the
electric and magnetic dipole moments and the driving term coming from the
different quadrupole moments coupled to themselves and the dipole moments.

We point out, since we have been keeping terms only to second order, that we
have left out a natural cubic term that comes from the mass loss equation.
If we write 
\[
P^{k}=Mv^{k}+........ 
\]
we see that 
\[
P^{k\prime }=Mv^{k\prime }+M^{\prime }v^{k}+....... 
\]
It is the term $M^{\prime }v^{k}$ that we have omitted. From Eq.(\ref{m'})
we have the cubic term that could have been included in the equations of
motion: 
\begin{equation}
P^{k\prime }=Mv^{k\prime }-v^{k}\{\frac{288c}{5G}%
(v_{R}^{ij}v_{R}^{ij}+v_{I}^{ij}v_{I}^{ij})+\frac{2q^{2}}{3c^{5}}
(v_{R}^{i\prime }v_{R}^{i\prime }+v_{I}^{i\prime }v_{I}^{i\prime })+\frac{
32q^{2}}{5c^{5}}(v_{R}^{ij\prime }v_{R}^{ij\prime }+v_{I}^{ij\prime
}v_{I}^{ij\prime })\}+.......  \label{cubic}
\end{equation}
Actually this cubic term, without the gravitational contribution, (the first
cubic term), is well known in classical electrodynamics\cite{Thir}

Summarizing what has been found from the dynamical equations we see; (a)
From the mass loss equations we have been able to identify or give the
physical meaning to the $\xi _{R}^{i}$, $\xi _{I}^{i},$ $\xi ^{ij},$ etc.,
in terms of different moments, (b) From the Kerr metric, we identified $%
Mc\xi _{I}^{i}$ with the spin vector and from the charged Kerr metric (and
also from the mass loss equation) we associate $q\xi _{I}^{i}$ with the
magnetic dipole moment. This leads immediately to the Dirac value of the
gyromagnetic ratio, $g=2$, (c) Finally we have the momentum loss equation
and the equations of motion for the `center of mass', $\xi _{R}^{i}$
determining the motion of an gravitational/electromagnetic rocket. Without
any attempt at model building we have automatically found the well-known
radiation reaction force - classically leading to the unpleasant run-away
behavior - but now with damping forces. Though it is not at all clear if the
damping terms are sufficient to stop the run-away behavior, nevertheless $if$
there was an existence proof for the exact type II solutions to
asymptotically approach the Kerr-Newman metric that would be a strong
indication for the suppression of the run-away affects. On the other hand it
might well be that the type II metrics with a Maxwell field are in fact
unstable.

\section{Conclusion}

In this work we have returned to the study of the important class of
Algebraically Special Type II metrics and tried to give them physical
content. Many attempts over the past forty or so years have been made to
solve them - with very few exact solutions - most work being on
approximations. Existence theorems for the pure vacuum case (in the
Robinson-Trautman subclass\cite{Piotr}) were given and much approximate work
was done on them but there appears to have been little progress on the
Maxwell version. There have been several attempts\cite{Drey,Tafel}, via
approximations, to relate, at least, the pure RT metrics to the more general
class of asymptotically flat Bondi metrics where one has an understanding of
asymptotic energy and momentum. However, to our knowledge, little or nothing
along these lines was done for the RTM equations\cite{private}.

Here, with the help of the UCF, many of the ideas have been unified with a
possible resolution of the physical content of these metrics. We have even
given the relationship of these metrics to the Bondi metrics, via the UCF.

Though later in this section we will speculate on some of the physical
significance of our observations, we first summarize our general
construction.

In an arbitrary asymptotically flat space-times, we started with families of 
\textit{regular} asymptotic shear-free null geodesic congruences, described
by its (stereographic) angle-field $L(u,\zeta ,\overline{\zeta })$ at $\frak{%
\ I}^{+},$ that are generated by arbitrary complex curves in H-space. This
function, $L,$ and the associated curve are then completely determined by
the requirement that the $l=1$ harmonic of the asymptotic Weyl tensor
component, $\psi _{_{1}},$ should vanish. This leads to a unique complex
world-line and UCF, $u=X(\tau ,\zeta ,\overline{\zeta }),$ which then is
expanded in spherical harmonics. This general procedure was then applied to
the analysis of the type II metrics - using severe approximations. When the
harmonic coefficients of $X(\tau ,\zeta ,\overline{\zeta })$ were used in
the Bondi mass aspect (in particular, the Bondi energy-momentum four-vector)
and the Bondi evolution equation, we were able to assigned to them `natural
appearing' physical meaning.

\begin{itemize}
\item  First, by separating the real and imaginary parts, we could identify
the $l=0,1,$ harmonics and their derivatives, as the center of mass position
and velocity vectors for the real parts and the specific spin-angular
momentum for the imaginary part. Then, from the mass aspect, we saw how the
Bondi energy and momentum took on a kinematic meaning in terms of these
kinematic variables.

\item  From the Kerr metric and the charged Kerr metric, we identified $%
Mc\xi _{I}^{i}$ with the spin vector. From the mass loss equation, we
associate $q(\xi _{R}^{i},\xi _{I}^{i})$ with the electric and magnetic
dipole moments. This leads immediately to the Dirac value of the
gyromagnetic ratio, $g=2$. Other multiple moments extracted from $X(\tau
,\zeta ,\overline{\zeta }),$ were identified via the energy loss theorem.

\item  Finally, using the kinematic expression for the momentum, in the
Bondi momentum loss equation we obtained equations of motion for the `center
of mass', determining the motion of what could be identified as a
gravitational/electromagnetic rocket.
\end{itemize}

Without any attempt at model building we automatically had the well-known
radiation reaction force - which classically leads to the unpleasant
run-away behavior but now it contains damping forces. Though it is not at
all clear if the damping terms are sufficient to stop the run-away behavior,
nevertheless \textit{if} there was an existence proof for the exact type II
solutions to asymptotically approach the Kerr-Newman metric that would be a
strong indication for the suppression of the run-away affects\cite{private}.

Several questions concerning these results immediately come to mind.

\begin{itemize}
\item  In what space is this motion taking place?

\item  In what sense could it be said that this motion is observable?

\item  Are there any possible consequences that can be derived from these
results?
\end{itemize}

There are tentative answers to each. As we pointed out earlier, the motion
can be thought of as taking place in either the space of the complex
Poincar\'{e} translation group or in H-space with the velocity norm coming
from either the Minkowski or H-space metric. Though neither of these spaces
are observable directly, in principle - but almost certainly not in practice
- the motion could be deduced by observing the special unique class of
asymptotically shear-free null rays, and then studying the $l=0,1$
coefficients of the harmonic decomposition of their angular distribution. In
other words the complex curve is there (if these shear-free rays could be
found) but in a strange form.

The real physical meaning of these results certainly is not at all clear.
They do contain basic well-known classical equations: Newton's 2nd Law, rest
and kinetic energy, the usual expression for momentum, the electromagnetic
dipole radiation formula, the Dirac value of the gyromagnetic ratio, the
classical radiation reaction force, the gravitational quadrupole formula. In
other words there is a strong suggestion that they do have meaning. If that
were the case then they should be taken seriously and one should look for
new physical results and predictions. For example, in Eq.(\ref{M****}) we
can see other contributions to the mass/energy (in addition to rest energy
and kinetic energy) coming from the spin and from the electric charge.
Though very small they are being predicted.

Another interesting possibility arises from the radiation reaction force.
This term in the equations of motion has been a source of considerable
difficulty for classical electromagnetic theory\cite{Thir,LL}. Almost all
attempts to study the classical motion of charged particles interacting with
their own field has led to the unphysical run-away behavior. The usual
answer to this problem has been to invoke quantum or quantum field theory%
\cite{Thir}. Our suggestion or conjecture is that when our procedure is
applied to any asymptotically flat Einstein-Maxwell space-time (with the two
world-lines coinciding), assuming that the space-time is well behaved and
approaches Reissner-Nordstrom or charged Kerr, \textit{then there will be a
radiation reaction term in the center of mass equations of motion whose
influence }\textbf{will be suppressed }\textit{by gravitational affects}.
There already are tentative results in this direction. This calculation is
being planned for the near future. To do the calculation \textit{without}
the assumption of the coincidence of the two world-lines, though planned for
the future, will be quite difficult.

As a final comment we mention that though there has been much mathematical
work done on CR structures and GR (see Sec. IV, Remark 3), it appears to us
that this is the first time that a CR structure has made relevant contact
with physical issues.

\section{Acknowledgments}

This material is based upon work (partially) supported by the National
Science Foundation under Grant No. PHY-0244513. Any opinions, findings, and
conclusions or recommendations expressed in this material are those of the
authors and do not necessarily reflect the views of the National Science.
E.T.N. thanks the NSF for this support. G.S.O. acknowledges the financial
support from Sistema Nacional de Investigadores (SNI-M\'{e}xico). C.K.
thanks CONICET and SECYTUNC for support. I.G.S.S. thanks CONACyT for support
via a scholarship.


\begin{thebibliography}{99}
\bibitem{private}  Recent work by Kozameh, Reula and Kreiss indicates that
the type II Robinson-Trautman metrics with a Maxwell field do in general
possess run-away behavior.
\end{thebibliography}
\end{document}